\documentclass[11pt]{article}
\usepackage[utf8]{inputenc}
\usepackage[english]{babel}
\usepackage{amsmath,amsfonts,amsthm,graphicx}
\usepackage{natbib}
\usepackage{algorithm,algpseudocode,float}

\def\bm{{\mathbf m}}

\def\b1{{\mathbf 1}}

\usepackage[font=small]{caption}

\newtheorem{theorem}{Theorem}

\usepackage{titlesec}

\titleformat*{\section}{\normalfont\fontsize{14}{17}\bfseries}
\titleformat*{\subsection}{\normalfont\fontsize{12}{15}\selectfont}

\usepackage{bm}
\usepackage{enumerate}
\providecommand{\abs}[1]{\lvert#1\rvert}
\providecommand{\norm}[1]{\lVert#1\rVert}

\usepackage{soul}

\date{}

\begin{document}

\title{\LARGE A computational validation for nonparametric assessment of spatial trends}\normalsize
\author{Andrea Meil\'an-Vila \\
Universidade da Coru\~{n}a\thanks{%
Research group MODES, Department of Mathematics, Faculty of Computer Science, Universidade da Coru\~na, Campus de Elvi\~na s/n, 15071,
A Coru\~na, Spain}
\and
Rub\'en Fern\'andez-Casal\\
Universidade da Coru\~{n}a\footnotemark[1]
\and Rosa M. Crujeiras \\
Universidade de Santiago de Compostela\thanks{Department of Statistics, Mathematical Analysis and Optimization, Faculty of Mathematics, Universidade de Santiago de Compostela, R\'ua Lope G\'omez de Marzoa s/n,
	15782, Santiago de Compostela, Spain}
\and %
Mario Francisco-Fern\'andez\\
Universidade da Coru\~{n}a\footnotemark[1]
}
\maketitle


\begin{abstract}
The analysis of continuously spatially varying processes usually considers two sources of variation, namely, the large-scale variation collected by the trend of the process, and the small-scale variation. Parametric trend models on latitude and longitude are easy to fit and to interpret. However, the use of simple parametric models for characterizing spatially varying processes may lead to misspecification problems if the model is not appropriate. Recently, \cite{meilan2019goodness} proposed a goodness-of-fit test based on an $L_2$-distance for assessing a parametric trend model with correlated errors, under random design, comparing a parametric and a nonparametric trend estimators.   The present work aims to provide a detailed computational analysis of the behavior of this approach using different bootstrap algorithms for calibration, under a fixed-design geostatistical framework.  Asymptotic results for the test  are provided and an  extensive simulation study, considering complexities that usually arise in geostatistics, is carried out to illustrate the performance of the proposal.

\end{abstract}	
\textit{Keywords:} {Parametric spatial trends, Bootstrap algorithm, Nonparametric fit, Goodness-of-fit test, Bias correction}

\section{Introduction}
\label{sec:intro}
Continuously varying spatial processes are usually described through the analysis of their trend and their dependence structure. The trend component captures the large-scale variability of the process, usually modeled by parametric functions on latitude and longitude (the so-called trend-surface models), and possibly altitude \citep[see][]{cressie1993statistics}. For a proper estimation of parametric trends with no distributional assumption, the dependence structure of the process (although not being of primary interest) must be accounted for, usually employing iterative least squares procedures or maximum likelihood approaches under Gaussian and stationary assumptions \citep[see, for instance,][]{diggle2010geostatistical,cressie1993statistics}.

However, the consideration of an inadequate parametric trend model may lead to wrong conclusions on the process behavior. In a regression setting, to prevent for misspecification  of the regression function, testing procedures for assessing a certain parametric regression model have been proposed, considering independent errors. For instance, the proposals by \cite{hardle1993comparing,alcala1999goodness,opsomer2010finding} and \cite{li2005using} are based on the comparison of a parametric and a nonparametric estimator using an $L_2$-distance.  Following this idea,  \cite{meilan2019goodness} introduced a testing procedure to check if the trend function of a spatial process belongs to a certain class of parametric models.
The authors considered a random design multivariate regression model with spatially correlated errors and used the local linear estimator as a  nonparametric fit.
In the present paper, focused on a geostatistical framework with fixed-design observations, a thorough analysis of the behavior of a similar test considering different trend models, sample sizes and dependence patterns, is  provided. In this case, for simplicity, the Nadaraya--Watson estimator is employed as a nonparametric fit.

The proposed test statistic shows a slow rate of convergence to its asymptotic distribution, motivating the use of resampling methods to approximate its distribution under the parametric null hypothesis. 
It should be noted that, in order to mimic the process behavior under the null hypothesis, not only the parametric form of the trend has to be considered, but also the spatial dependence of the data, which has to be recovered from a single realization of the spatial process, under stationarity conditions.  In the presence of spatial correlation, resampling methods may not be accurate enough for mimicking the  spatial dependence structure under the null hypothesis from a single realization of the process. This is the reason why a thorough analysis of the impact of the spatial dependence configuration in the distribution approximation is required and provided in this work.

Traditional resampling procedures for test calibration designed for independent
data should not be used for spatial processes, as they do not account for the correlation
structure. One of the  aims of this paper is to present and analyze three different proposals for test calibration which take the dependence of the data into account: a parametric residual bootstrap (PB), a nonparametric residual bootstrap (NPB) and a bias-corrected nonparametric bootstrap (CNPB). Parametric bootstrap procedures, following the ideas introduced by \cite{solow1985bootstrapping}, are a usual strategy in geostatistics, since they directly involve the dependence structure \cite[see, for example,][]{olea2011generalized}. The PB approach consists in using in the bootstrap algorithm the residuals obtained from the parametric fit and, from these residuals, estimating parametrically the spatial dependence structure. If the trend function indeed belongs to the parametric family considered in the null hypothesis, then the residuals obtained with this approach will be \emph{similar} to the theoretical errors, and it is expected that the PB method will have a good performance. A possible drawback of this procedure is the misspecification of the
parametric model selected for the dependence estimation, however this issue could be avoided by using  a
nonparametric estimator instead. However, this resampling approach relies on the wrong assumption that the variability of the residuals is the same as the one of the theoretical errors.  In the NPB method, to increase the power of the test, residuals are obtained from the nonparametric fit \citep[see][]{gonzalez1993testing}. Moreover, the dependence structure is estimated without considering parametric assumptions. It is clear that the NPB resamplig method can avoid the misspecification problem  both for the trend and the dependence. However, no matter the method used to remove the trend, either
parametric  or nonparametric, the direct use of residuals gives rise to biased variogram estimates,
especially at large lags \cite[see][Section 3.4.3]{cressie1993statistics}. To solve this problem, the CNPB approach is a modification of the NPB method, but including a bias-corrected algorithm for the dependence estimation \citep[see][]{fernandez2014nonparametric,castillo2019nonparametric}.

This paper is organized as follows. In Section \ref{sec:inference}, the parametric and nonparametric methods used to estimate the spatial trend employed in our testing procedure are briefly described. The $L_2$-test statistic measuring the discrepancy between both fits as well as its asymptotic distribution are also included in Section \ref{sec:inference}. A detailed description of the calibration algorithms considered is given in Section \ref{sec:practice}. In addition, an exhaustive simulation study to assess the performance of the test, when the PB, NPB and CNPB resampling approaches are used, is presented is Section \ref{sec:simulation}. Finally, Section \ref{sec:discussion} includes some discussion and further considerations.

\section{Inference for spatial trends}
\label{sec:inference}

A research question that is usually addressed by spatial modeling is the estimation of a surface/map describing the trend of the process, or the prediction of the variable of interest at certain unobserved locations. To tackle these problems,  traditional  approaches in geostatistics consist in assuming (generally simple) parametric models for the trend and, then, to reconstruct the whole trend surface using parametric techniques and make predictions by spatial interpolation methods, such as kriging \citep[see][Section 3]{cressie1993statistics}. However, in some situations, the complex interactions between the possible factors affecting the variable of interest make it difficult to write down a simple parametric model for its trend over a large geographic region. In the absence of such a model, representing the trend as a smooth spatial function, establishing the relation between the process values and the location coordinates (latitude, longitude and maybe altitude in a three-dimensional setting), provides a useful first step to characterize important features of the variable of interest, or can be helpful in the development of more complete models including additional significant covariates.

Consider a real-valued spatial process $\{Z(\mathbf s),\;\mathbf s\in D \subset\mathbb R^d\}$, observed at fixed locations $\{\mathbf s_1,\ldots,\mathbf s_n\}$. From a model-based perspective  \citep[see][]{ribeiro2007model}, the spatial process can be assumed to be decomposed as:
\begin{equation}
Z_i=m(\mathbf s_i)+\varepsilon_i,\quad i=1,\ldots,n,
\label{eq:model}
\end{equation}
being $Z_i=Z(\textbf{s}_i)$, with $i=1,\ldots,n$, a realization of the process at a collection of locations within the observation domain. The trend of the process is given by $m$, which is an unknown (but smooth) function modeling the  expectation of the process, and $\varepsilon_i$ denotes the error at location $\mathbf s_i$, for $i=1,\ldots,n$, so these values can be viewed as a realization of a spatially varying error process. In order to estimate the trend in (\ref{eq:model}) from a single realization of the process, stationary conditions must be assumed. Usually, the error process is supposed to be zero-mean with covariance structure satisfying:
\begin{equation}\label{eq:cov}
\mbox{Cov}(\varepsilon_i,\varepsilon_j)=\sigma^2\rho_n(\mathbf s_i-\mathbf s_j), \quad i,j=1,\ldots,n,
\end{equation}
with $\sigma^2$ being the variance of the process and $\rho_n$ a continuous stationary correlation function satisfying $\rho_n(0)=1$, $\rho_n(\mathbf{s})=\rho_n(-\mathbf{s})$, and $\abs{\rho_n(\mathbf{s})}\le1$, $\forall \mathbf{s}$. The subscript $n$ in $\rho_n$ allows the correlation function to shrink as $n\to\infty$.   In a spatial context, the dependence structure is typically characterized through the variogram function, $\gamma_n$, which satisfies that $\gamma_n(\mathbf{s})=\sigma^2(1-\rho_n(\mathbf{s}))$, $\forall\mathbf{s}\in \mathbb{R}^d$.
For simplicity, the subscript $n$ will be sometimes omitted. In the previous expression for the covariance of the errors (\ref{eq:cov}), it is supposed that the nugget effect is zero. Otherwise, the variance of the errors is written as the sum of two terms, $\mbox{Var}(\varepsilon)=\sigma^2= c_0+c_e$, the nugget effect ($c_0$) and the partial sill ($c_e$), and $\mbox{Cov}(\varepsilon_i,\varepsilon_j)=c_e\rho_n(\mathbf{s}_i-\mathbf{s}_j),$ if $i\neq j$. In what follows, the covariance matrix of the errors is denoted by $\Sigma$, being $\Sigma(i,j)=\mbox{Cov}(\varepsilon_i,\varepsilon_j)$ its $(i,j)$-entry. For the sake of simplicity, no nugget  is considered in the theoretical result given in Section \ref{sec:testing}. However, its effect is analyzed in the simulation study presented in Section \ref{sec:simulation}.

In model (\ref{eq:model}), the trend function $m$ can be characterized using parametric or nonparametric models. Parametric models are easy  to compute and allow for a direct interpretation of the parameter values (e.g. variation of the process along latitude and longitude). On the other hand, nonparametric models also provide a global view of the behavior of the large-scale variability process. Their flexibility allows to model complex relations beyond a parametric form. Therefore, a question of interest in spatial modeling is focused on characterizing the large-scale variability of the process $Z$, checking if the trend function belongs to a parametric family by solving the following testing problem:
\begin{equation}
H_0:\;m\in \mathcal{M}_{\mathbf\beta}=\{m_{\mathbf\beta},\;\mathbf\beta\in\mathcal{B}\}\quad\mbox{vs.}\quad H_a:\;m\notin \mathcal{M}_{\mathbf\beta},
\label{eq:test}
\end{equation}
with $\mathcal{B}\subset\mathbb{R}^p$ a compact set , and $p$ denotes the dimension of the parameter space $\mathcal{B}$. 

A test statistic  to address (\ref{eq:test}) is proposed and studied in this paper. Following similar ideas to those in \citet{meilan2019goodness}, the proposed test consists in using a nonparametric fit as a pilot estimator to assess if a certain parametric family is suitable for fitting the observed data, comparing with an $L_2$-distance the nonparametric fit with a parametric one.

In the following section, the parametric and the nonparametric estimators of the spatial trend $m$, in model (\ref{eq:model}), used in our $L_2$-test statistic will be described. Subsequently, the asymptotic distribution of this test will be derived and its empirical performance will be analyzed in a comprehensive simulation study, under different spatial dependence scenarios.

\subsection{Spatial trend estimation}
\label{sec:estimation}

Spatial trend estimation in (\ref{eq:model}) can be performed parametrically by different methods, being least squares and maximum likelihood approaches the most frequently used \citep{ribeiro2007model}. Next, we briefly describe the parametric least squares trend estimator used in our test statistic. On the other hand, nonparametric methods can also be employed for this task. Among the different alternatives, the multivariate Nadaraya--Watson estimator will be applied in the goodness-of-fit test proposed. This nonparametric trend estimator is also formulated and discussed below in our context of interest.

\subsubsection*{Parametric estimation}
\label{sec:estimation_par}

As pointed out previously, the goodness-of-fit test proposed in this paper makes use of a parametric estimator of the trend function. As it will be remarked in the next section, the test statistic can be applied considering any parametric estimator of $m$ satisfying a $\sqrt{n}$-consistency property. Specifically, if $m_{{{\mathbf{\beta}}}_0}$ denotes the ``true'' regression function under the null hypothesis, and $m_{\hat{{\mathbf{\beta}}}}$ the corresponding parametric estimator, it is needed that the difference  $m_{\hat{{\mathbf{\beta}}}}(\mathbf{s})-m_{{{\mathbf{\beta}}}_0}(\mathbf{s})=O_p(n^{-1/2})$ uniformly in $\mathbf{s}$.
A suitable parametric estimator satisfying this property is, for example, the one considered by \cite{rosa_csda} for nonlinear trends. 
The steps of the parametric estimation method employed for the practical application of the test are the following:
\begin{enumerate}
	\item Obtain an initial estimator of $\mathbf{\beta}$ by  least squares regression:
	\begin{equation}
	\tilde{\mathbf{\beta}}=\mbox{arg}\min_{{\mathbf{\beta}}}(\mathbf Z-\mathbf m_{\mathbf{\beta}})'(\mathbf Z-\mathbf m_{\mathbf{\beta}}), 
	\label{eq:OLS}
	\end{equation}
	where $Z=(Z_1,\ldots,Z_n)'$ and  $\mathbf m_{\mathbf\beta}=(m_{\mathbf\beta}(\mathbf s_1),\ldots,m_{\mathbf\beta}(\mathbf s_n))'$.
	\item Using the residuals obtained with the estimation in (\ref{eq:OLS}), $\tilde{\varepsilon}_i=Z_i-m_{\tilde{\mathbf{\beta}}}(\mathbf{s}_i)$, $i=1,\dots,n$, estimate the covariance matrix  of the errors, $\tilde\Sigma$.  
	\item Update the regression parameter estimates, introducing the estimated covariance matrix $\tilde\Sigma$ in the least squares minimization problem:
	\begin{equation}
	\hat{\mathbf{\beta}}=\mbox{arg}\min_{\mathbf{\beta}}(\mathbf Z-\mathbf m_{\mathbf{\beta}})'\tilde\Sigma^{-1}(\mathbf Z-\mathbf m_{\mathbf{\beta}}).
	\label{eq:IGLS}  
	\end{equation}
	Finally, take $m_{\hat{\mathbf{\beta}}}$ as the parametric estimator for the regression function.
\end{enumerate}

Covariance matrix estimation in Step 2  could be carried out using different methods. Firstly, using a parametric approach and assuming that the variogram belongs to a valid parametric family $\{2\gamma_{\bm{\theta}},\;\bm{\theta}\in\bm{\Theta}\subset\mathbb{R}^q\}$,  a parameter estimate $\hat{\bm{\theta}}$ of $\bm{\theta}$ can be obtained.
Fo\-llo\-wing a classical approach, $\bm{\theta}$ could be estimated by fitting the parametric model considered for the variogram to a set of empirical semivariogram estimates, computed using the residuals $\tilde{\varepsilon}_i$, applying the weighted  least squares method \citep{cressie1985fitting}.  
With this parametric approximation, the variance-covariance matrix of the errors can be denoted by $\Sigma_{\bm{\theta}}$, and replacing $\bm{\theta}$ by $\hat{\bm{\theta}}$, a parametric estimation of $\Sigma_{\bm{\theta}}$ (denoted by ${\Sigma}_{\hat{\bm{\theta}}}$) can be obtained.

On the other hand, instead of using a parametric approach, flexible nonparametric estimators can be employed to approximate the dependence structure, avoiding misspecification problems. For instance, an estimate of the variogram of the
residuals could be obtained as follows. First, compute a nonparametric pilot variogram estimator \citep{hall1994properties}. A first attempt could be to use the empirical semivariogram estimator.  Nevertheless, in practice, semivariogram models fitted to the empirical variogram could be unsatisfactory. For instance, the assumption of
isotropy (or geometric anisotropy) could be not appropriate \citep{fernandez2003flexible}. Therefore, it would  be  desirable to have  models
with enough flexibility. Nonparametric kernel semivariogram estimators could be used instead, producing significantly better results than those obtained with the empirical estimator \citep{fernandez2003space}.  However, these estimators do not necessarily satisfy the conditionally negative definiteness property of a valid semivariogram.  For that reason, a valid model should be fitted to the nonparametric pilot estimates. For example, a flexible Shapiro--Botha variogram approach \citep{shapiro1991variogram} could be employed at this step. The combination of the Shapiro--Botha approach  with nonparametric kernel semivariogram pilot estimation provides an efficient variogram estimator which can be used to estimate the corresponding covariance matrix.

\subsubsection*{Nonparametric estimation}
\label{sec:estimation_npar}

Kernel methods can also be used to estimate the trend function $m$ in model (\ref{eq:model}), providing more flexible approaches than the parametric fits. In this work, a multivariate Nadaraya--Watson estimator \citep{hardle1997multivariate,liu2001kernel} is considered. For a certain location $\mathbf s \in D$, this estimator is given by:
\begin{equation}
\hat{m}^{NW}_{\mathbf{H}}(\mathbf s)=\dfrac{\sum_{i=1}^{n}K_{\mathbf H}(\mathbf s_i-\mathbf s)Z_i}{\sum_{i=1}^{n}K_{\mathbf H}(\mathbf s_i-\mathbf s)},
\label{eq:NW}
\end{equation}
where $K_\mathbf{H}(\mathbf{s})=\abs{\mathbf{H}}^{-1}K(\mathbf{H}^{-1}\mathbf{s})$ is the rescaled version of a multivariate  kernel function $K$ and  $\mathbf{H}$ is a $d\times d$ symmetric positive definite matrix.   The kernel function $K$ can be obtained as the product of univariate kernels \citep[see, for example,][]{wand1994kernel,fan1996local}.  The bandwidth matrix $\mathbf{H}$ controls the shape and the size
of the local neighborhood used to estimate $m$ at a location $\mathbf{s}$, and its selection plays an important role in the estimation process.  In the presence of spatially correlated errors,  traditional
data-driven bandwidth selection methods, such as cross-validation and generalized
cross-validation, fail to provide good bandwidth values.  
Asymptotic results for (\ref{eq:NW}) as well as the proposal of different bandwidth selection methods under the assumption of spatially correlated errors can be found in \cite{liu2001kernel}, extending the results for independent data given in \cite{ruppert1994multivariate}.

The estimator given in (\ref{eq:NW}) can be seen as a particular case of a wider class of nonparametric estimators, the so-called local polynomial estimators, assuming that the polynomial degree is equal to zero (local constant). Since this work aims to provide a deeper analysis of the practical performance of a version of the test studied in \cite{meilan2019goodness}, the local constant fit was chosen given its reduced computational cost compared with other nonparametric approaches.

\subsection{Trend model assessment}
\label{sec:testing}

Following the ideas by \cite{hardle1993comparing} and \cite{alcala1999goodness}, \cite{meilan2019goodness} addressed the testing problem (\ref{eq:test}) constructing a weighted $L_2$-test statistic, comparing a parametric and a nonparametric regression estimates. The authors considered a model like (\ref{eq:model}), but assuming a random design. As parametric fit, they used the least squares estimator described in Section \ref{sec:estimation_par}, estimating the covariance matrix of  the errors in Step 2 of that algorithm using a parametric approach. As a nonparametric fit, they employed the local linear estimator. In the present paper, we consider  a geostatistical fixed-design and the Nadaraya--Watson estimator introduced in (\ref{eq:NW}). Specifically, the test statistic  is given by:

\begin{equation}
T_n=n|\mathbf H|^{1/2}\int_{D}^{}\left(\hat{m}^{NW}_{\mathbf H}(\mathbf s)-\hat{m}^{NW}_{\mathbf H,\hat{\mathbf\beta}}(\mathbf s)\right)^2w(\mathbf s)d\mathbf s,
\label{eq:statistic}
\end{equation}
where $w$ is a weight function that helps in mitigating possible boundary effects and $\hat{m}^{NW}_{\mathbf H,\hat{\mathbf\beta}}$ is a smooth version of ${m}_{\hat{\mathbf\beta}}$, which is defined by:
\begin{equation}
\hat{m}^{NW}_{\mathbf H,\hat{\mathbf\beta}}(\mathbf s)=\dfrac{\sum_{i=1}^n K_{\mathbf H}\left(\mathbf s_i-\mathbf s\right)m_{\hat{\mathbf\beta}}(\mathbf s_i)}{\sum_{i=1}^nK_{\mathbf H}\left(\mathbf s_i-\mathbf s\right)}.
\label{eq:NWpar}
\end{equation}

\begin{figure}[h!]\centering
	\includegraphics[width=6.5cm]{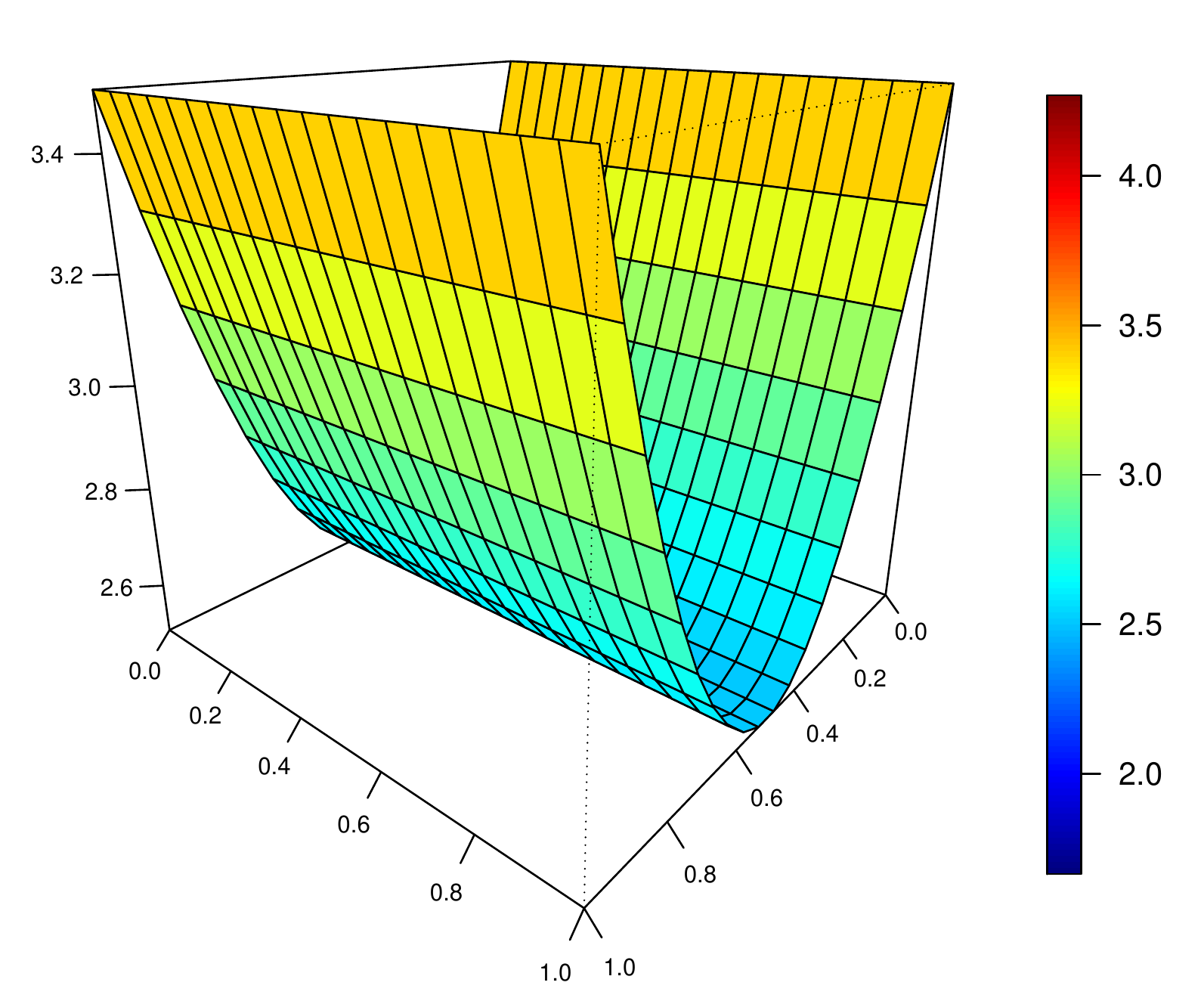}\hspace{0.4cm}
	\includegraphics[width=6.5cm]{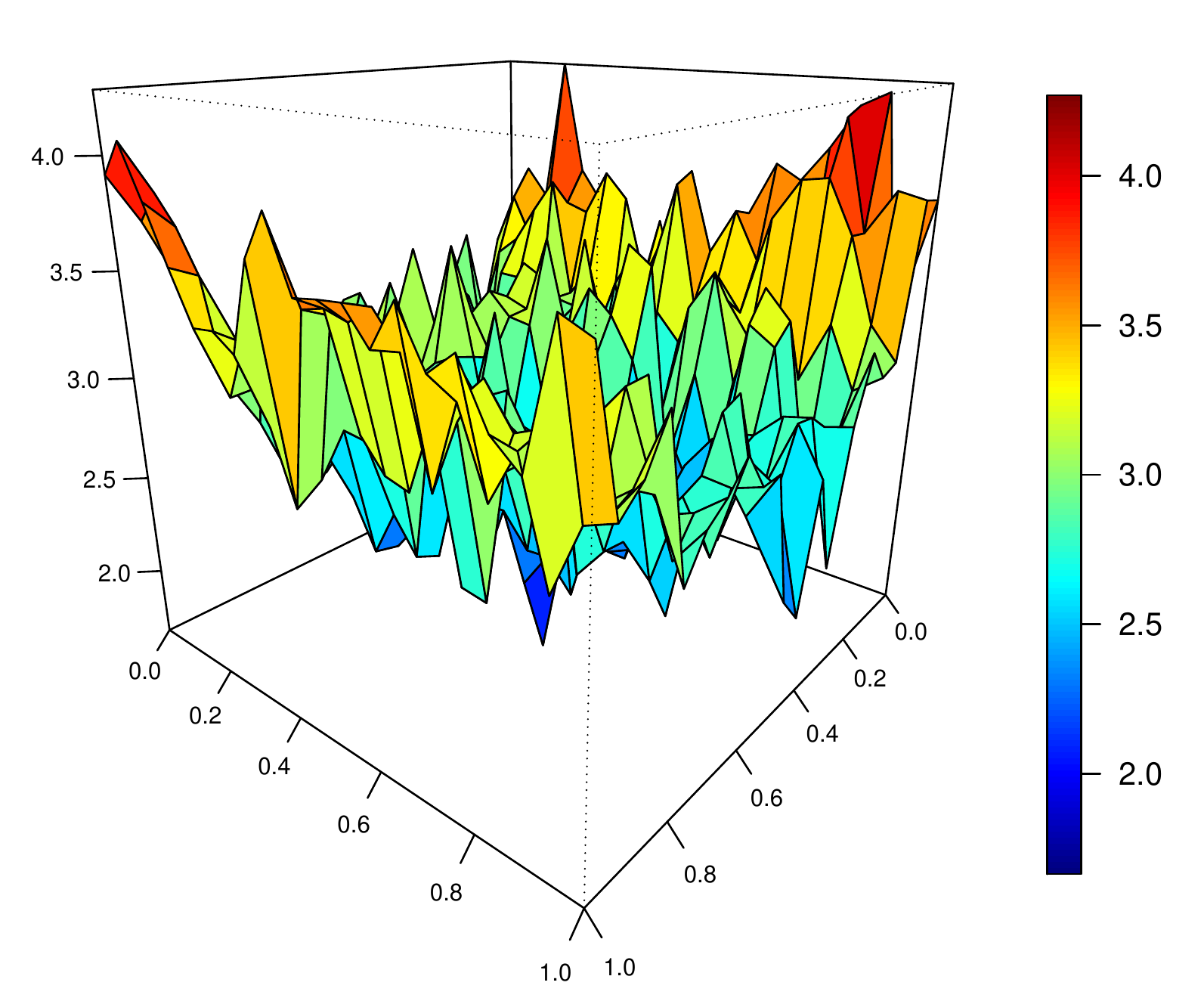}\vspace{0.4cm}	
	\includegraphics[width=6.5cm]{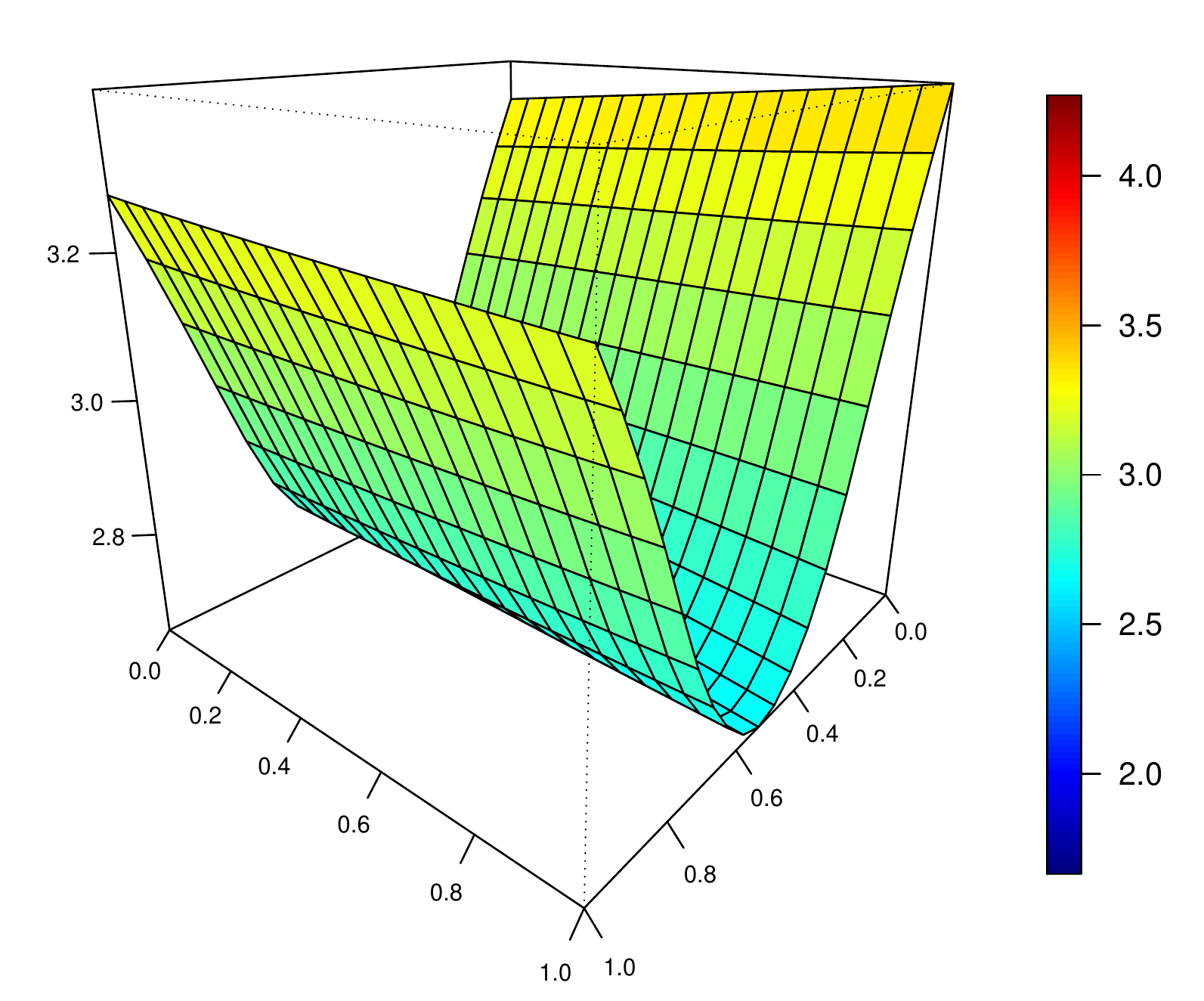}\hspace{0.4cm}
	\includegraphics[width=6.5cm]{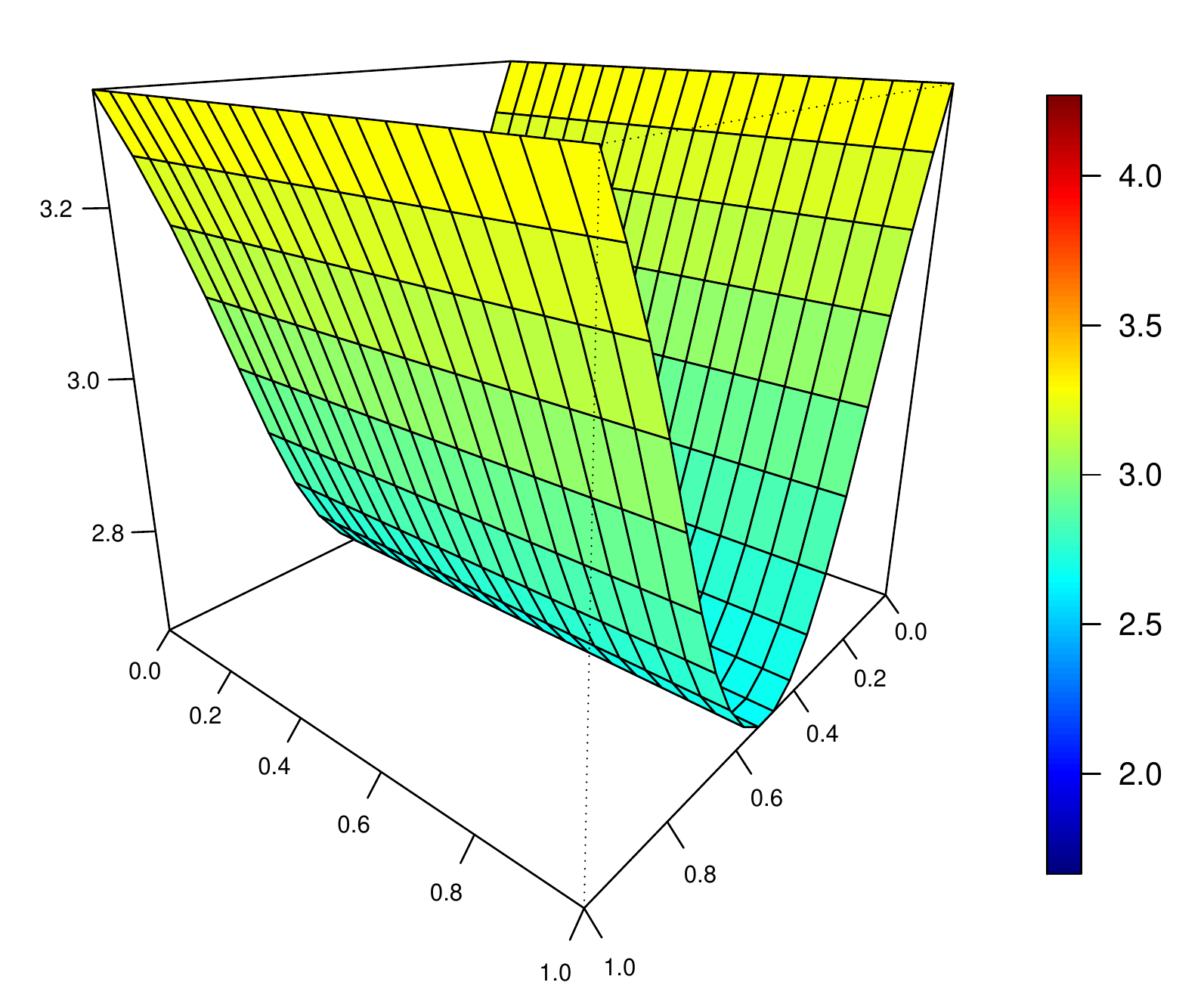}
	\caption{Spatial trend (top left),  spatial process (top right), Nadaraya--Watson trend estimator (bottom left) and smooth version of the parametric fit (bottom right). Sample of size $n=400$ generated on a bidimensional regular grid in the unit square, following model (\ref{eq:model}), with trend function $m(\mathbf{s})=2.5+4 (s_1-0.5)^3$, $\mathbf{s}=(s_1,s_2)$, and exponential  covariance structure  with $\sigma^2=0.16$, $c_0=0.04$ and  $a_e=0.6$}
	\label{fits}
\end{figure}

If the null hypothesis is true, then the parametric and nonparametric estimators in (\ref{eq:statistic}) will tend to be \emph{similar} and the value of $T_n$ will be \emph{small}. Conversely, if the null hypothesis is false, major differences between  both fits will be expected,  and therefore, the value of $T_n$ will be  large. So, $H_0$ will be  rejected if the distance between both fits exceeds a critical value. For example, as a visual illustration of the performance of the test, suppose that a sample of size $n=400$ is generated  following model (\ref{eq:model}), with trend function (\ref{trend_sim1}), with $c=0$, and  random errors  normally distributed with zero mean and  covariance function (\ref{eq:exp_cov}), with values $\sigma^2=0.16$, $c_0=0.04$ and  $a_e=0.6$. If we want to test if $m\in\{\beta_0+\beta_1(s_1-0.5)^3, \beta_0,\beta_1\in\mathbb{R}\}$ using the test statistic given in (\ref{eq:statistic}), both $\hat{m}^{NW}_{\mathbf H}$ and $\hat{m}^{NW}_{\mathbf H,\hat{\mathbf\beta}}$ fits must be computed. Figure \ref{fits} shows the theoretical spatial trend (top left panel), the simulated observations of the spatial process (top right panel), the Nadaraya--Watson trend estimation (bottom left panel) and the smooth version of the parametric fit  (bottom right panel). A multiplicative triweight kernel and the optimal bandwidth obtained by minimizing the Mean Average Squared Error (MASE) of the Nadaraya--Watson  estimator \citep[see][p. 288]{francisco2005smoothing} are considered for $\hat{m}^{NW}_{\mathbf H}$ and $\hat{m}^{NW}_{\mathbf H,\hat{\mathbf\beta}}$. In this case, from a visual comparison, one may argue that given that both estimates at bottom left and right panels are very similar,  the value of the test statistic $T_n$ is \emph{small},  and consequently, there is no evidences against the assumption of parametric trend $m_{\mathbf{\beta}}(\mathbf{s})=\beta_0+\beta_1 (s_1-0.5)^3$. However, apart from getting some insight to what might occur when using exploratory methods, in order to formally test the model using $T_n$ given in (\ref{eq:statistic}), it is essential to approximate the distribution of the test statistic under the null hypothesis. The types of model deviations that can be detected by the test given in (\ref{eq:statistic}) are of the form $m(\mathbf{s})=m_{{\mathbf{\beta}}_0}(\mathbf{s})+ c_n g(\mathbf{s})$, where $c_n$ is a sequence, such that $c_n\to0$ and $g$ is a deterministic function collecting the deviation direction from the null model. Specifically, it is assumed that the function $g$ is bounded  and $c_n=n^{-1/2}\abs{\mathbf{H}}^{-1/4}$. In particular, it contains the null hypothesis for $g(\mathbf{s})=0$.  A result providing the asymptotic distribution of the test statistic $T_n$  is given below. The following assumptions are required:

\begin{enumerate}[{(A}1)]
\item The regression function $m$ is  twice continuously differentiable.
	\item The weight function $w$ is continuously differentiable. 
	
	\item For the correlation function $\rho_n$, there exist  $\rho_{M}$ and $\rho_{c}$ such that
	$n\int \abs{\rho_n(\mathbf s)}d\mathbf s<\rho_{M}$ and $\lim_{n\to\infty}n\int {\rho_n(\mathbf s)}d\mathbf s=\rho_{c}.$
	For any sequence $\epsilon_n>0$ satisfying $n^{1/d}\epsilon_n\to\infty$,
	$$n\int_{\norm{\mathbf s}\ge\epsilon_n} \abs{\rho_n(\mathbf s)}d\mathbf s\to 0\quad \text{as}\quad n\to\infty.$$
	
	\item For any $i$, $j$, $k$, $l$,
	$$
	\mbox{Cov}({\varepsilon}_i{\varepsilon}_j,{\varepsilon}_k{\varepsilon}_l )=\mbox{Cov}({\varepsilon}_i,{\varepsilon}_k)\mbox{Cov}({\varepsilon}_j,{\varepsilon}_l)+\mbox{Cov}({\varepsilon}_i,{\varepsilon}_l)\mbox{Cov}({\varepsilon}_j,{\varepsilon}_k).$$
	\item The errors are a geometrically strong mixing sequence with mean zero and $\mathbb{E}\abs{\varepsilon(\mathbf s)}^r<\infty$ for all $r> 4$.
	\item The kernel $K$ is a spherically symmetric density function, twice continuously differentiable and with compact support (for simplicity with a nonzero value only if $\norm{\mathbf{u}}\le 1$). Moreover, $\int \mathbf{u}\mathbf{u}' K(\mathbf{u})d\mathbf{u}=\mu_2(K)\mathbf{I}_d$, where $\mu_2(K)$ is a constant real value different from zero and $\mathbf{I}_d$ is the $d\times d$ identity matrix.
	
	\item $K$ is Lipschitz continuous. That is, there exists $\mathfrak{L}>0$, such that 
	$$\abs{K(\mathbf s_1)-K(\mathbf s_2)}\le \mathfrak{L}\norm{\mathbf s_1-\mathbf s_2}, \quad \forall \mathbf s_1,\mathbf s_2\in D.$$

	\item The bandwidth matrix $\mathbf{H}$ is symmetric and positive definite, with $\mathbf{H}\to 0$ and  $n\abs{\mathbf{H}}\lambda^2_{\min}(\mathbf{H})\to\infty$, when $n\to\infty$. 
	The ratio $\lambda_{{\max}}(\mathbf{H})/\lambda_{{\min}}(\mathbf{H})$ is bounded previous, where $\lambda_{{\max}}(\mathbf{H})$ and $\lambda_{{\min}}(\mathbf{H})$ are the maximum and minimum eigenvalues of $\mathbf{H}$, respectively.
	
\end{enumerate}

Assumption (A3) implies that the correlation function depends on $n$, and  the integral $\int \abs{\rho_n(\mathbf s)}d\mathbf s$ should vanish  as $n\to\infty$. The vanishing speed should not be slower than $O(n^{-1})$. This assumption also  implies that the integral of $\abs{\rho_n(\mathbf s)}$  is  dominated by the values of $\rho_n(\mathbf s)$ near to the origin. Hence, the correlation
is short-range and decreases as $n\to\infty$. This can be considered as a
case of increasing-domain spatial asymptotics \citep[see][]{cressie1993statistics}, since this setup can 
be transformed to one in which the correlation function $\rho_{n}$ is fixed with respect to the sample
size, but the support  $D$
for $\mathbf s$ expands. The current setup with fixed domain $D$ and shrinking $\rho_n$ is more natural to consider when the primary purpose of the estimation is a fixed regression function $m$ defined over a spatial domain, not the correlation function itself. Two examples of commonly used correlation functions that satisfy the conditions of assumption (A3) are the exponential and rational quadratic models  \citep[see][]{cressie1993statistics}. Assumption (A4)  is satisfied, for example, for Gaussian errors. (A5) is needed to apply the central limit theorem for reduced $U$-statistics under dependence given by \cite{kim2013central}. In assumption (A8), $\mathbf{H}\to 0$ means that every entry of $\mathbf{H}$ goes to $0$. Since $\mathbf{H}$ is symmetric and positive definite, $\mathbf{H}\to 0$ is equivalent to $\lambda_{{\max}}(\mathbf{H})\to 0$. $\abs{\mathbf{H}}$ is a quantity of order $O(\lambda_{{\max}}^d(\mathbf{H}))$ given that $\abs{\mathbf{H}}$ is equal to the product of all eigenvalues of $\mathbf{H}$. 

Regarding the parametric estimator,  the assumption of being a $\sqrt{n}$-con\-sis\-tent estimator is required. This is guaranteed if the least squares estimator $m_{\hat{{\mathbf{\beta}}}}$ des\-cri\-bed in Section \ref{sec:estimation} is used in the statistic (\ref{eq:statistic}). A different parametric estimator of the trend function
could be used as long as $\sqrt{n}$-con\-sis\-tency property is fulfilled.  

The following theorem provides the asymptotic distribution of the proposed test statistic $T_n$ given in (\ref{eq:statistic}). The proof of the result can be found in the final Appendix.

\begin{theorem}\label{theorem}
	Under Assumptions \textnormal{(A1)--(A8)}, and  if $0<V<\infty$, it can be proved that
	\begin{equation*}
	V^{-1/2}(T_n-b_{0\mathbf{H}}-b_{1\mathbf{H}})\to_{\mathcal{L}} N(0,1) \text{ as } n\to\infty
	\end{equation*}	where $\to_{\mathcal{L}}$ denotes convergence in distribution, with
	\begin{eqnarray*}
		b_{0\mathbf{H}}&=& |\mathbf{H}|^{-1/2}\sigma^2K^{(2)}(\mathbf 0)\left(\int w(\mathbf s)d\mathbf s+\rho_{c}\int {w(\mathbf s)}d\mathbf s\right),\\
		b_{1\mathbf{H}}&=& \int \left(K_{\mathbf{H}}\ast g(\mathbf s)\right)^2w(\mathbf s)d\mathbf s,\\
	\end{eqnarray*}
	and
	\begin{eqnarray*}
		V&=&\sigma^4 K^{(4)}(0)\int_{}^{} w^2(\mathbf s)d\mathbf s\left(1+\rho_{c}+2\rho^2_c\right),
	\end{eqnarray*}
	where $K^{(j)}$ denotes the $j$-times convolution product of $K$ with itself.
\end{theorem}

\section{Testing proposal in practice}
\label{sec:practice}
Notice that the asymptotic distribution of the test obtained in Theorem 1, as in other nonparametric testing procedures \citep[see, for example,][]{hardle1993comparing}, may not be sufficiently precise to approximate the test statistic distribution under the null hypothesis in practice,  for small or moderate sample sizes.  Given the slow rate of convergence, to obtain an accurate approximation of the asymptotic distribution of the test, it would be necessary to have a large sample size, which is not always the case for geostatistical data. Moreover,  the limit distribution of the test statistic depends on unknown quantities that must be estimated.  
This is a common issue in smoothing-based tests, as already noted by \cite{gonzalez2013updated}, which is usually overcome using resampling methods, specifically, employing  bootstrap algorithms that try to mimic the data structure under the null hypothesis.

In what follows, a detailed description of the different bootstrap proposals designed to perform the calibration of the test (namely PB, NPB and CNPB) will be presented. The main difference between the proposals is how the resampling residuals (required for mimicking the dependence structure) are computed. In PB, the residuals are obtained from the parametric trend estimator. 
Alternately, in NPB, the residuals are obtained from the nonparametric trend estimator \citep[see][]{gonzalez1993testing}. In this way, the error variability could be reproduced consistently both under the null and the alternative hypotheses, increasing the power of the test. Finally, the CNPB procedure is a modification of the NPB, where the residuals are also obtained from the nonparametric trend estimator, but, additionally, the variability is estimated with an iterative algorithm to correct the bias due to the use of the residuals \citep{fernandez2014nonparametric}. 

In order to describe the PB, NPB and CNPB resampling approaches, a generic bootstrap algorithm is firstly introduced. In what follows, no matter the method used, either
parametric or nonparametric, $\hat{m}$ and $\hat{\Sigma}$ denote the trend and the covariance matrix estimates, respectively.    

\begin{algorithm*}[h!]
	\caption{}
	\begin{algorithmic}
		\State 1. Compute a parametric or a  nonparametric trend estimator (described in Section \ref{sec:estimation_par}), namely $\hat{m}(\mathbf s_i)$, $i=1,\ldots,n$, depending if a parametric (PB) or a  nonparametric (NPB or CNPB) bootstrap procedure is employed.
		\State 2.	Obtain an estimated variance-covariance matrix   $\hat{\Sigma}$ of the residuals   $\hat{\mathbf\varepsilon}=(\hat{\varepsilon}_1,\ldots,\hat{\varepsilon}_n)'$,   where  $\hat{\varepsilon}_i=Z_i-\hat{m}(\mathbf s_i)$, $i=1,\ldots,n$.
		\State 3. Find the matrix $L$, such that  $\hat{\Sigma}=LL'$, using Cholesky decomposition.
		\State 4. Compute the \emph{independent} variables, $\mathbf{e}=(e_1,\ldots,e_n)'$, given by $\mathbf{e}=L^{-1}\mathbf{\hat{\varepsilon}}$.
		\State 5. The previous \emph{independent} variables are centered and an \emph{independent} bootstrap sample of size $n$, denoted by $\mathbf{e}^*=(e^*_1,\ldots,e^*_n)$, is obtained.
		\State 6. The bootstrap errors  $\mathbf{\varepsilon}^*=(\varepsilon^*_1,\ldots,\varepsilon^*_n)$ are computed as $\mathbf {\varepsilon}^*=L\mathbf{e}^*$, and the bootstrap samples are  $Z^*_i=m_{\hat{\mathbf{\beta}}}(\mathbf s_i)+{\varepsilon}^*_i$, being $m_{\hat{\mathbf{\beta}}}(\mathbf s_i)$  the parametric trend estimator. 
		\State 7.  Using the bootstrap sample $\{Z^*_i, i=1,\ldots,n\}$, the bootstrap test statistic $T_n^*$ is computed as in (\ref{eq:statistic}).
		\State 8. Repeat Steps 4-7 a large number of times $B$.
	\end{algorithmic}
\end{algorithm*}

The empirical distribution of the $B$ boostrap test statistics can be employed to approximate  the finite sample distribution of the  test statistic $T_n$ under the null hypothesis. Thus,  denoting by $\{T_{n,1}^*,\cdots,T_{n,B}^*\}$ the sample of the  $B$ bootstrap test statistics,  and defining its $(1-\alpha)$ quantile $t_\alpha^*$,  the null hypothesis in (\ref{eq:test}) will be rejected if $T_n>t_\alpha^*$. Additionally, the $p$-value of the test statistic can be approximated by:
$$p\mbox{-value}=\dfrac{1}{B}\sum_{b=1}^B\mathbb{I}_{\{T_{n,b}^*>T_n\}}.$$ 

Some steps of the mentioned algorithm are discussed below for PB, NPB and NCPB methods. The main differences between the procedures are highlighted.

\subsection{Parametric residual bootstrap (PB)}
The PB extends to the case of spatial trends the parametric residual bootstrap discussed in \cite{vilar1996bootstrap}. In Step 1 of the previous algorithm, the trend is estimated parametrically, employing  the iterative least squares estimator described in Section \ref{sec:estimation_par}.  In Step 2, from the parametric residuals, the covariance matrix is also computed using a parametric approach (see Section \ref{sec:estimation_par} for further details).  Notice that if the trend and the semivariogram  belong to the assumed  parametric families, then this procedure should provide good results. However,  a drawback of this procedure is the misspecification problem that may affect the trend and variance estimation. Moreover, as it was pointed out in the Introduction, the direct use of the residuals   introduces a bias in the estimation of the variability  of the process in Step 2.

\subsection{Nonparametric residual bootstrap (NPB)}

The NPB  tries to avoid the misspecification problems mentioned in the previous section by using more flexible trend and dependence estimation methods than those employed in PB. In Step 1 of the bootstrap algorithm, to increase the power of the test, following \citet{gonzalez1993testing},  the Nadaraya--Watson estimator given in (\ref{eq:NW}) is employed. In addition, in Step 2,  a flexible procedure is considered to estimate the covariance matrix. The  Shapiro--Botha variogram approach \citep{shapiro1991variogram}, combined  with a nonparametric kernel semivariogram pilot estimation provides an efficient variogram estimator, which is used to approximate the corresponding covariance matrix. For more details see Section \ref{sec:estimation_par}.

\subsection{Corrected nonparametric residual bootstrap (CNPB)}
As it was pointed out before, no matter the method used to remove the trend in Step 2, either
parametric  or nonparametric, the direct use of the residuals  in the variogram estimation introduces a bias in the approximation of the process variability. The CNPB procedure is a modification of the previous NPB approach, considering a procedure to correct  the resulting bias  in the nonparametric estimator of the variogram.  In the geostatistical framework, more accurate results have been obtained using this technique \citep{castillo2019nonparametric}. 
Specifically, the following adjustments are performed  in the previous generic bootstrap algorithm.  In Step 1, the trend is estimated
using the Nadaraya--Watson estimator given in (\ref{eq:NW}), whereas in Step 2, from the corresponding nonparametric residuals,  the dependence structure is estimated nonparametrically with an iterative algorithm to correct
the bias   \citep{fernandez2014nonparametric}. Moreover, two additional steps are included  after Step 2 and 3, which are denoted by $2^\ast$ and $3^\ast$:
\begin{enumerate}
	\item[$2^\ast.$]  Obtain a bias-corrected estimate of the variogram, using the residuals obtained from the nonparametric fit  \cite[see][for an exhaustive description of the algorithm]{fernandez2014nonparametric} and calculate the corresponding (estimated) covariance matrix $\tilde{\Sigma}$ of the errors. 
	\item [$3^\ast.$] Find the matrix $\tilde{L}$, such that  $\tilde{\Sigma}=\tilde{L}\tilde{L}'$, using Cholesky decomposition. $\tilde{\Sigma}=\tilde{L}\tilde{L}'$.
\end{enumerate}

In this situation, for the CNPB method, Step 6 in the algorithm needs to be modified as follows:
\begin{enumerate}[6.]	\item The bootstrap errors  $\mathbf{\varepsilon}^*=(\varepsilon^*_1,\ldots,\varepsilon^*_n)$ are $\mathbf {\varepsilon}^*=\tilde{L}\mathbf{e}^*$, and the bootstrap samples are  $Z^*_i=m_{\hat{\mathbf{\beta}}}(\mathbf s_i)+{\varepsilon}^*_i$, where $m_{\hat{\mathbf{\beta}}}(\mathbf s_i)$ was computed using the procedure described in Section \ref{sec:estimation_par}.\end{enumerate}

\section{Simulation study}
\label{sec:simulation}
In this section, the practical performance of the proposed test statistic is analyzed through a simulation study comparing the different bootstrap procedures described in Section \ref{sec:practice}.

Two regression models are considered. First,  the parametric trend family $\mathcal{M}_{1,\mathbf\beta}=\{\beta_0+\beta_1(s_1-0.5)^3, \beta_0,\beta_1\in\mathbb{R}\}
$  is assumed for the null hypothesis. In this case, the theoretical trend functions are given by:
\begin{equation}
m_1(\mathbf{s})=2.5+4 (s_1-0.5)^3+c\sin(2\pi s_2), \quad \mathbf s=(s_1,s_2).
\label{trend_sim1}
\end{equation}

The second parametric trend family considered for the null hypothesis   is $\mathcal{M}_{2,\mathbf\beta}=\{\beta_0+\beta_1 \cos(\pi s_1) , \beta_0,\beta_1\in\mathbb{R}\}$,  and the trend functions are:
\begin{equation}
m_2(\mathbf{s})=1+2\cos(\pi s_1)+c\sin(2\pi s_2), \quad \mathbf s=(s_1,s_2).
\label{trend_sim2}
\end{equation}

In both cases, the parameter $c$ controls whether the null ($c=0$) or the alternative ($c\neq 0$) hypotheses  hold. For different values of this parameter, 500 samples of sizes $n$ ($n=100,225$ and $400$) are generated on a bidimensional regular grid in the unit square, following model (\ref{eq:model}), with regression functions (\ref{trend_sim1}) or (\ref{trend_sim2}), and 
random errors $\varepsilon_i$ normally distributed with  zero mean and isotropic exponential covariogram:  
\begin{equation}
\label{eq:exp_cov}
\mbox{Cov}({\varepsilon}_i,{\varepsilon}_j)=   
c_e\left[\exp(-\norm{\mathbf s_i-\mathbf s_j}/a_e)\right],
\quad \norm{\mathbf s_i-\mathbf s_j}\neq 0,\end{equation}
where $c_e$ is the partial sill and $a_e$ the practical range, while the variance of the errors (also called the sill) is $\sigma^2=c_0+c_e$, being $c_0$ the nugget effect.  Different degrees of spatial dependence were
studied, considering values of $a_e=0.3, 0.6$ and $0.9$,
$\sigma^2=0.16, 0.32$ and $0.32$, and nugget values of $0\%, 25\%$ and $50\%$ of $\sigma^2$.

The behavior of the test statistic given in (\ref{eq:statistic})  was analyzed in the different scenarios. The parametric fit used to construct (\ref{eq:statistic}) was computed using the iterative least squares procedure described in Section \ref{sec:estimation_par}. The nonparametric fit was obtained using the multivariate Nadaraya--Watson estimator, given in (\ref{eq:NW}), with a multiplicative triweight kernel.  The bandwidth selection 
problem was addressed by employing the same procedure as that used in \cite{hardle1993comparing}, \cite{alcala1999goodness}, or \cite{opsomer2010finding}, among others, analyzing the performance of the test statistic $T_n$ in (\ref{eq:statistic}) for a range of bandwidths. This allows to check how sensitive the results are to variations in $\mathbf{H}$. Initially, to simplify the calculations,  the bandwidth matrix was restricted to a diagonal matrix with both equal elements (scalar matrix), $\mathbf{H}=\text{diag}(h,h)$, and different  values of $h$ in the interval $[0.25, 1.50]$ were chosen.
The weight function employed in (\ref{eq:statistic}) to avoid the possible boundary effect \citep{gonzalez1993testing} was $w(\textbf{s})=\mathbb{I}_{\{\textbf{s}\in[1/\sqrt{n},1-1/\sqrt{n}]\times[1/\sqrt{n},1-1/\sqrt{n}]\}}$, where $\mathbb{I}_{\{\cdot\}}$ denotes the indicator function.

The bootstrap  procedures described in Section \ref{sec:practice} were applied using  $B=500$ replications.  In the nonparametric residual bootstrap procedures, NPB and CNPB, the multivariate Nadaraya--Watson estimator was computed in Step 1  using the optimal bandwidth
that minimizes the MASE. Similar results were obtained when the corrected generalized cross-validation (CGCV) bandwidth is employed \citep{francisco2005smoothing}. However, the use of the MASE bandwidth matrix  reduces the computing time and avoids the effect of the bandwidth selection for the trend estimation on the results. Regarding the variogram, the (uncorrected) variogram
estimates  and the bias-corrected version were computed
on a regular grid up to the $55\%$ of the largest sample distance. In this case, the bandwidth matrices were selected applying the
cross-validation relative squared error criterion. 

\begin{figure*}[h!]
	\centering
	\includegraphics[width=1\textwidth]{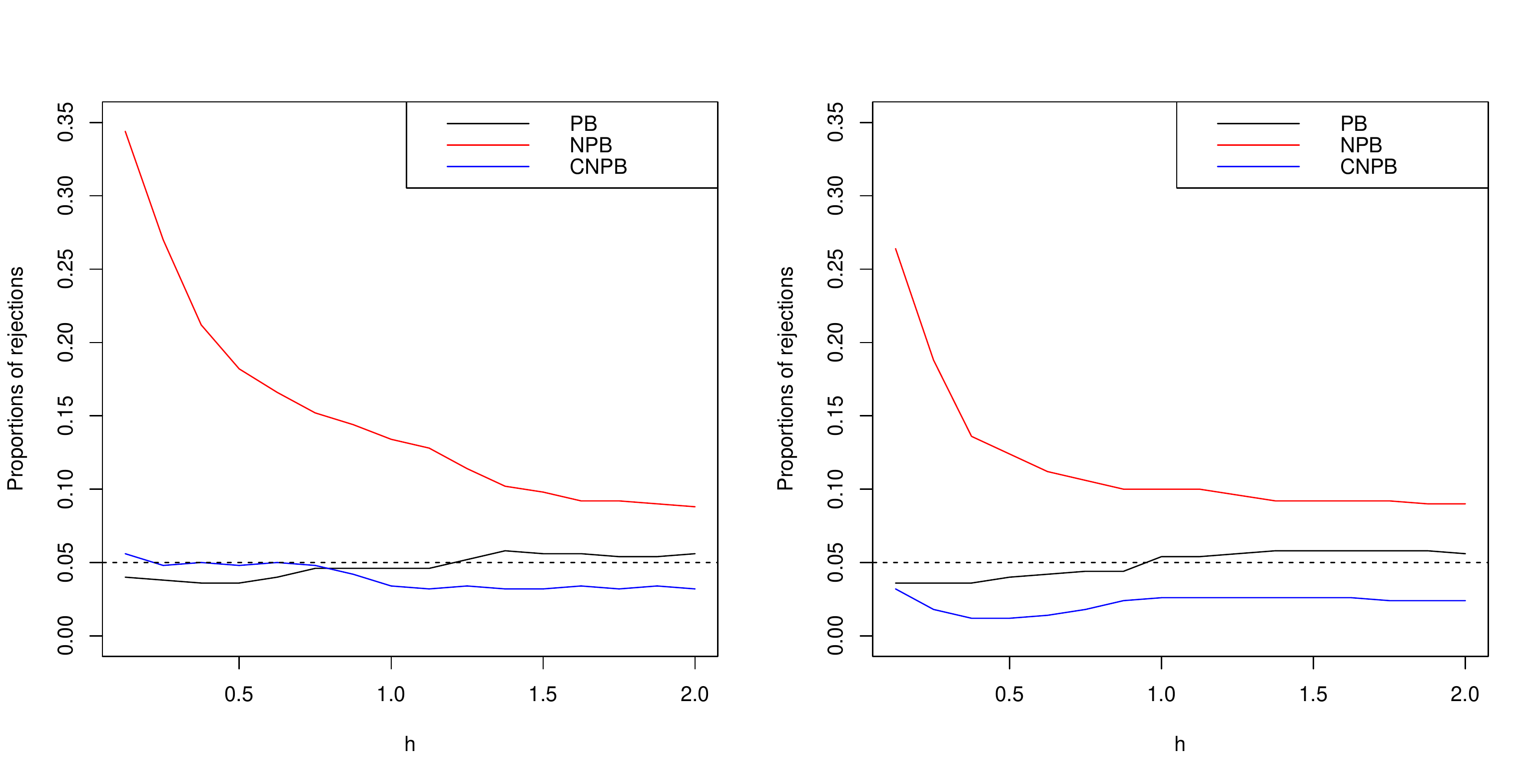}
	\caption{Proportions of rejections under the  null hypothesis $(c=0)$ for trend functions (\ref{trend_sim1}) (left panel) and  (\ref{trend_sim2}) (right panel). Model parameters: $c_0=0.04$, $\sigma^2=0.16$, $a_e=0.6$ and $n=400$. Significance level: $\alpha=0.05$}
	\label{fig::simu_trends}
\end{figure*}

\begin{table}[h!]
	\caption{Proportions of rejections of the null hypothesis for the parametric family  $\mathcal{M}_{1,\mathbf\beta}$ with different sample sizes. Model parameters: $c_0=0.04$, $\sigma^2=0.16$, $a_e=0.6$. Significance level: $\alpha=0.05$}
	\label{table::simu_n}
	\begin{tabular}{lllllllll}
		\hline
		$c$&			$n$&\text{Method}&$h=0.25$&$h=0.50$&$h=0.75$&$h=1.00$&$h=1.25$&$h=1.50$\\
		\hline
		0&	100 & PB & 0.056 & 0.042 & 0.054 & 0.066 & 0.070 & 0.068 \\ 
		&	& NPB & 0.340 & 0.216 & 0.170 & 0.142 & 0.102 & 0.088 \\ 
		&	& CNPB & 0.064 & 0.050 & 0.040 & 0.028 & 0.018 & 0.018 \\ 
		
		&	225 & PB & 0.070 & 0.068 & 0.060 & 0.070 & 0.082 & 0.082 \\ 
		&	& NPB & 0.268 & 0.192 & 0.176 & 0.144 & 0.116 & 0.108 \\ 
		&	& CNPB & 0.078 & 0.058 & 0.046 & 0.042 & 0.030 & 0.028 \\ 
		
		&	400 & PB & 0.038 & 0.036 & 0.046 & 0.046 & 0.052 & 0.056 \\ 
		&	& NPB & 0.270 & 0.182 & 0.152 & 0.134 & 0.114 & 0.098 \\ 
		&	& CNPB & 0.048 & 0.048 & 0.048 & 0.034 & 0.034 & 0.032 \\\hline
		0.5&	100 & PB & 0.000 & 0.002 & 0.006 & 0.022 & 0.026 & 0.036 \\ 
		&& NPB & 1.000 & 0.996 & 0.978 & 0.960 & 0.898 & 0.818 \\ 
		&& CNPB & 0.740 & 0.574 & 0.380 & 0.198 & 0.076 & 0.050 \\ 
		&225 & PB & 0.018 & 0.012 & 0.020 & 0.046 & 0.066 & 0.074 \\ 
		&& NPB & 1.000 & 0.994 & 0.976 & 0.938 & 0.846 & 0.734 \\ 
		&& CNPB & 0.692 & 0.550 & 0.398 & 0.218 & 0.102 & 0.056 \\ 
		&400 & PB & 0.004 & 0.004 & 0.016 & 0.036 & 0.058 & 0.070 \\ 
		&& NPB & 1.000 & 1.000 & 0.984 & 0.940 & 0.852 & 0.716 \\ 
		&& CNPB & 0.384 & 0.316 & 0.208 & 0.076 & 0.034 & 0.028 \\

		\hline
		1&		100 & PB & 0.056 & 0.010 & 0.018 & 0.038 & 0.074 & 0.110 \\ 
		&	& NPB & 1.000 & 1.000 & 1.000 & 1.000 & 0.994 & 0.982 \\ 
		&	& CNPB & 0.996 & 0.972 & 0.894 & 0.584 & 0.294 & 0.146 \\ 	
		&	225 & PB & 0.002 & 0.002 & 0.010 & 0.030 & 0.078 & 0.098 \\ 
		&	& NPB & 1.000 & 1.000 & 1.000 & 1.000 & 0.994 & 0.962 \\ 
		&	& CNPB & 0.990 & 0.938 & 0.848 & 0.564 & 0.296 & 0.126 \\ 	
		&	400 & PB & 0.000 & 0.000 & 0.000 & 0.024 & 0.064 & 0.098 \\ 
		&	& NPB & 1.000 & 1.000 & 1.000 & 1.000 & 1.000 & 0.968 \\ 
		&	& CNPB & 0.990 & 0.944 & 0.824 & 0.578 & 0.304 & 0.170 \\ 
		\hline
	\end{tabular}
\end{table}

Proportion of rejections (under the  null hypothesis, $c=0$) for several values of $h$ are plotted in Fig.~\ref{fig::simu_trends}. Left panel of Fig.~\ref{fig::simu_trends} shows the results for the trend (\ref{trend_sim1}) and right panel for the trend (\ref{trend_sim2}). In this simulation, a significance level of $\alpha=0.05$, and values of $c_0=0.04$, $\sigma^2=0.16$, $a_e=0.6$ and $n=400$ were considered.  Under the null hypothesis, the trend function belongs to the parametric family  and, as expected,  the  resampling procedure with a better performance is the parametric one (PB). On the other hand, although resampling methods following the rationale of NPB have provided good results in similar testing problems in other frameworks, for example, independent and univariate data \citep{gonzalez1993testing}, this is not the case in our geostatistical context. Using the NPB method,  the proportion of rejections of the null hypothesis is increased by the fact that the variability is  underestimated \citep[as noted by][]{fernandez2014nonparametric},  advising against the use of this procedure. Finally, the benefits of correcting the bias can  be observed, being the results for CNPB much better than those obtained with NPB.

The effect of the sample size as well as the spatial dependence, under the null and some alternative hypotheses,  is analyzed below. In the different scenarios considered, a comparison of the proposed bootstrap procedures (PB, NPB and CNPB) is presented. For the sake of brevity, only some  representative results employing the parametric family  $\mathcal{M}_{1,\mathbf\beta}$  are shown here. Similar conclusions were obtained when the  parametric family  $\mathcal{M}_{2,\mathbf\beta}$ was considered.

\subsection{Sample size effect}
In this section, the performance of the bootstrap procedures is analyzed for different sample sizes, under the null hypothesis and several alternatives. Proportions of rejections of the null hypothesis, for a significance level $\alpha=0.05$, considering the parameters $c_0=0.04$, $\sigma^2=0.16$, $a_e=0.6$ in model (\ref{eq:exp_cov}),  and different sample sizes,  are displayed in Table~\ref{table::simu_n}.    Under the null hypothesis ($c=0$), it can be observed that the test has a reasonable behavior when using PB and CNPB resampling methods. For both algorithms, the proportions of rejections are similar to the fixed significance level, although these proportions are slightly affected by the value of  $h$.  In fact, for CNPB, the proportions of rejections are smaller when the bandwidth value is larger. The opposite effect is observed when PB is employed. For alternative assumptions ($c=0.5$ and $c=1$), the performance of PB is really poor. A much better behavior is observed for CNPB.  A decreasing power is obtained when the value of $h$ increases. As expected, the power of the test becomes larger when the value of $c$ gets bigger.  Note that although it may seem that NPB presents a better behavior in terms of power, this is due to the underestimation of the variability, which induced really poor results under the null hypothesis.

\subsection{Range of dependence effect}
\begin{figure*}[h!]
	\centering
	\includegraphics[width=1\textwidth]{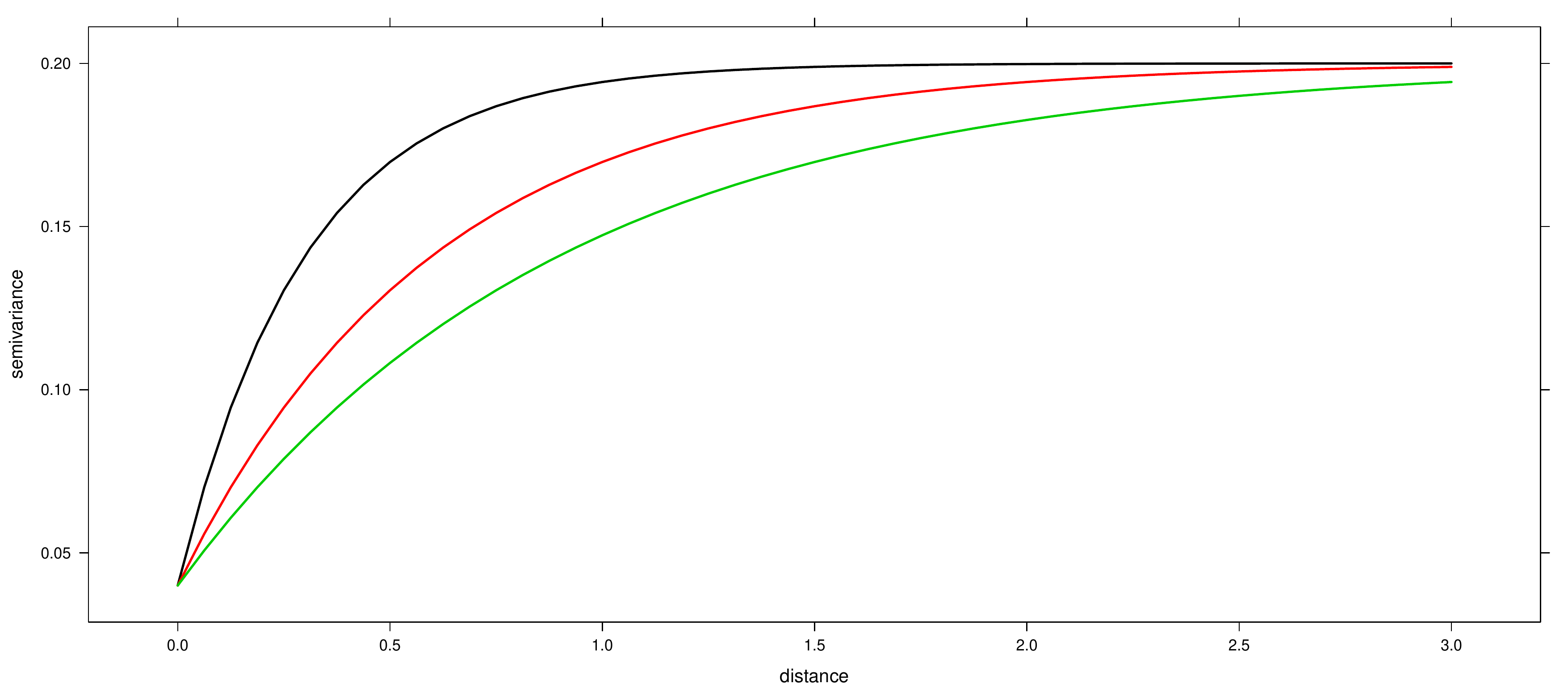}
	\caption{Exponential variogram models with $\sigma^2=0.16$ and $c_0=0.04$, for $a_e=0.3$ (black line), $a_e=0.6$ (red line) and $a_e=0.9$ (green line)}
	\label{variograms}
\end{figure*}

In this section, the performance of the different bootstrap procedures is analyzed for different spatial dependence degrees ($a_e=0.3$, $a_e=0.6$ and $a_e=0.9$).  Values of $n=400$, $\sigma^2=0.16$ and  $c_0=0.04$ are considered.  
Fig.~\ref{variograms} shows exponential variogram models  with $\sigma^2=0.16$ and $c_0=0.04$, for $a_e=0.3$ (black line), $a_e=0.6$ (red line) and $a_e=0.9$ (green line).
Table \ref{table::simu_ae} contains the proportions of rejections of the null hypothesis for $\alpha=0.05$. Notice that results for $a_e=0.6$ have already been shown in Table \ref{table::simu_n}, but for the sake of comparison they are also included in Table \ref{table::simu_ae}.  Again, it can be observed that CNPB provides good results for the null and the alternative hypotheses. As expected, for larger values of the practical range $a_e$, the bandwidth values providing an effective calibration of the test must also be larger. Regarding the PB approach, this resampling method  works properly under the null hypothesis (for appropriate values of the bandwidth parameters $h$), but its performance under the alternatives is very poor. On the other hand, although the NPB method has a very high power, the proportions of rejections under the null hypothesis are very large.
\begin{table}[h!]
	\caption{Proportions of rejections of the null hypothesis for the parametric family  $\mathcal{M}_{1,\mathbf\beta}$, with $\alpha=0.05$,  considering $c_0=0.04$, $\sigma^2=0.16$, $n=400$ and different range values}
	\label{table::simu_ae}
	\begin{tabular}{lllllllll}
		\hline
		$c$&			$a_e$&\text{Method}&$h=0.25$&$h=0.50$&$h=0.75$&$h=1.00$&$h=1.25$&$h=1.50$\\
		\hline
		0&	0.3 & PB & 0.056 & 0.050 & 0.040 & 0.044 & 0.044 & 0.050 \\ 
		&	& NPB & 0.158 & 0.090 & 0.076 & 0.062 & 0.050 & 0.052 \\ 
		&	& CNPB & 0.030 & 0.024 & 0.018 & 0.012 & 0.010 & 0.010 \\ 
		&	0.6 & PB & 0.038 & 0.036 & 0.046 & 0.046 & 0.052 & 0.056 \\ 
		&	& NPB & 0.270 & 0.182 & 0.152 & 0.134 & 0.114 & 0.098 \\ 
		&	& CNPB & 0.048 & 0.048 & 0.048 & 0.034 & 0.034 & 0.032 \\ 
		&	0.9 & PB & 0.002 & 0.004 & 0.018 & 0.030 & 0.040 & 0.042 \\ 
		&	& NPB & 0.328 & 0.266 & 0.232 & 0.186 & 0.162 & 0.144 \\ 
		&	& CNPB & 0.070 & 0.070 & 0.068 & 0.060 & 0.048 & 0.048 \\ 
		\hline
		0.5&0.3 & PB & 0.000 & 0.002 & 0.006 & 0.022 & 0.026 & 0.036 \\ 
		&& NPB & 1.000 & 1.000 & 0.998 & 0.976 & 0.878 & 0.716 \\ 
		&& CNPB & 0.346 & 0.270 & 0.170 & 0.042 & 0.018 & 0.016 \\ 
		&0.6 & PB & 0.004 & 0.004 & 0.016 & 0.038 & 0.058 & 0.066 \\ 
		&& NPB & 1.000 & 1.000 & 0.984 & 0.940 & 0.852 & 0.716 \\ 
		&& CNPB & 0.384 & 0.316 & 0.208 & 0.076 & 0.034 & 0.028 \\ 
		&0.9 & PB & 0.008 & 0.010 & 0.020 & 0.048 & 0.062 & 0.072 \\ 
		&& NPB & 1.000 & 0.996 & 0.982 & 0.954 & 0.880 & 0.750 \\ 
		& &CNPB & 0.526 & 0.432 & 0.284 & 0.146 & 0.074 & 0.052 \\ 
		\hline
		1&0.3 & PB & 0.000 & 0.000 & 0.000 & 0.010 & 0.050 & 0.062 \\ 
		&& NPB & 1.000 & 1.000 & 1.000 & 1.000 & 1.000 & 0.986 \\ 
		&& CNPB & 1.000 & 0.990 & 0.906 & 0.648 & 0.380 & 0.194 \\ 
		&0.6 & PB & 0.000 & 0.000 & 0.000 & 0.024 & 0.064 & 0.098 \\ 
		&& NPB & 1.000 & 1.000 & 1.000 & 1.000 & 1.000 & 0.968 \\ 
		&& CNPB & 0.990 & 0.944 & 0.824 & 0.578 & 0.304 & 0.170 \\ 
		&0.9 & PB & 0.000 & 0.000 & 0.008 & 0.030 & 0.080 & 0.110 \\ 
		&& NPB & 1.000 & 1.000 & 1.000 & 1.000 & 1.000 & 0.968 \\ 
		&& CNPB & 1.000 &0.962  &0.858  &0.662  &0.360  &0.206  \\ 
		\hline
	\end{tabular}
\end{table} 

\subsection{Nugget effect}
The performance of the proposed bootstrap procedures is now presented  for different values of the nugget,  $0\%, 25\%$ and $50\%$ of $\sigma^2$. Proportion of rejections of the null hypothesis are shown in Table \ref{table::simu_c0}, for  $\alpha=0.05$, considering $n=400$, $\sigma^2=0.16$ and $a_e=0.6$.  The best behavior is observed when CNPB is employed, showing a good performance for the null and the different alternative hypotheses. For larger values of the variogram at zero lag, smaller bandwidths must be taken to calibrate the test properly. On the other hand, no reliable results are obtained for NPB under the null and the alternative hypotheses. Finally, regarding PB, the power of the test is very small in all the scenarios considered.

\begin{table}[h!]
	\caption{Proportions of rejections of the null hypothesis for the parametric family  $\mathcal{M}_{1,\mathbf\beta}$, with $\alpha=0.05$,  considering $n=400$, $\sigma^2=0.16$, $a_e=0.6$ and different nugget effect values}
	\label{table::simu_c0}
	\begin{tabular}{lllllllll}
		\hline
		$c$&	$c_0$&\text{Method}&$h=0.25$&$h=0.50$&$h=0.75$&$h=1.00$&$h=1.25$&$h=1.50$\\
		\hline
		0&	$0\%$ & PB & 0.014 & 0.008 & 0.020 & 0.034 & 0.046 & 0.046 \\ 
		&	& NPB & 0.338 & 0.230 & 0.182 & 0.154 & 0.132 & 0.110 \\ 
		&	& CNPB &0.060  &0.058  &0.052  &0.044  &0.040  & 0.038 \\ 	
		&	$25\%$ & PB & 0.038 & 0.036 & 0.046 & 0.046 & 0.052 & 0.056 \\ 
		&	& NPB & 0.270 & 0.182 & 0.152 & 0.134 & 0.114 & 0.098 \\ 
		&	& CNPB & 0.048 & 0.048 & 0.048 & 0.034 & 0.034 & 0.032 \\	
		&	$50\%$ & PB & 0.050 & 0.048 & 0.042 & 0.048 & 0.052 & 0.056 \\ 
		&	& NPB & 0.254 & 0.172 & 0.144 & 0.116 & 0.092 & 0.082 \\ 
		&	& CNPB & 0.050 & 0.050 & 0.046 & 0.036 & 0.030 & 0.028 \\ \hline
		0.5&$0\%$ & PB & 0.006 & 0.014 & 0.034 & 0.052 & 0.062 & 0.068 \\ 
		&& NPB & 1.000 & 0.990 & 0.972 & 0.912 & 0.834 & 0.686 \\ 
		&& CNPB &0.418  &0.336  &0.240  &0.098  &0.050  &0.036  \\ 
		&$25\%$ & PB & 0.004 & 0.004 & 0.016 & 0.038 & 0.058 & 0.066 \\ 
		&& NPB & 1.000 & 1.000 & 0.984 & 0.940 & 0.852 & 0.716 \\ 
		&& CNPB & 0.640 & 0.534 & 0.384 & 0.206 & 0.096 & 0.058 \\ 
		&$50\%$ & PB & 0.000 & 0.002 & 0.010 & 0.026 & 0.048 & 0.056 \\ 
		&& NPB & 1.000 & 1.000 & 0.998 & 0.976 & 0.890 & 0.776 \\ 
		&& CNPB & 0.880 & 0.780 & 0.652 & 0.468 & 0.264 & 0.164 \\ 
		\hline 
		1&	$0\%$ & PB & 0.104 & 0.074 & 0.092 & 0.132 & 0.182 & 0.204 \\ 
		&	& NPB & 1.000 & 1.000 & 1.000 & 1.000 & 0.992 & 0.922 \\ 
		&	& CNPB &0.926  &0.830  &0.676  &0.378  &0.194  &0.080  \\ 
		&	$25\%$ & PB & 0.000 & 0.000 & 0.000 & 0.024 & 0.064 & 0.098 \\ 
		&	& NPB & 1.000 & 1.000 & 1.000 & 1.000 & 1.000 & 0.968 \\ 
		&	& CNPB & 0.990 & 0.944 & 0.824 & 0.578 & 0.304 & 0.170 \\ 
		&	$50\%$ & PB & 0.000 & 0.000 & 0.004 & 0.014 & 0.052 & 0.076 \\ 
		&	& NPB & 1.000 & 1.000 & 1.000 & 1.000 & 1.000 & 0.996 \\ 
		&	& CNPB & 1.000 & 1.000 & 0.970 & 0.840 & 0.576 & 0.378 \\ 
		\hline
	\end{tabular}
\end{table}

\subsection{More general bandwidth matrices} 

\begin{figure*}[h!]
	\includegraphics[width=1\textwidth]{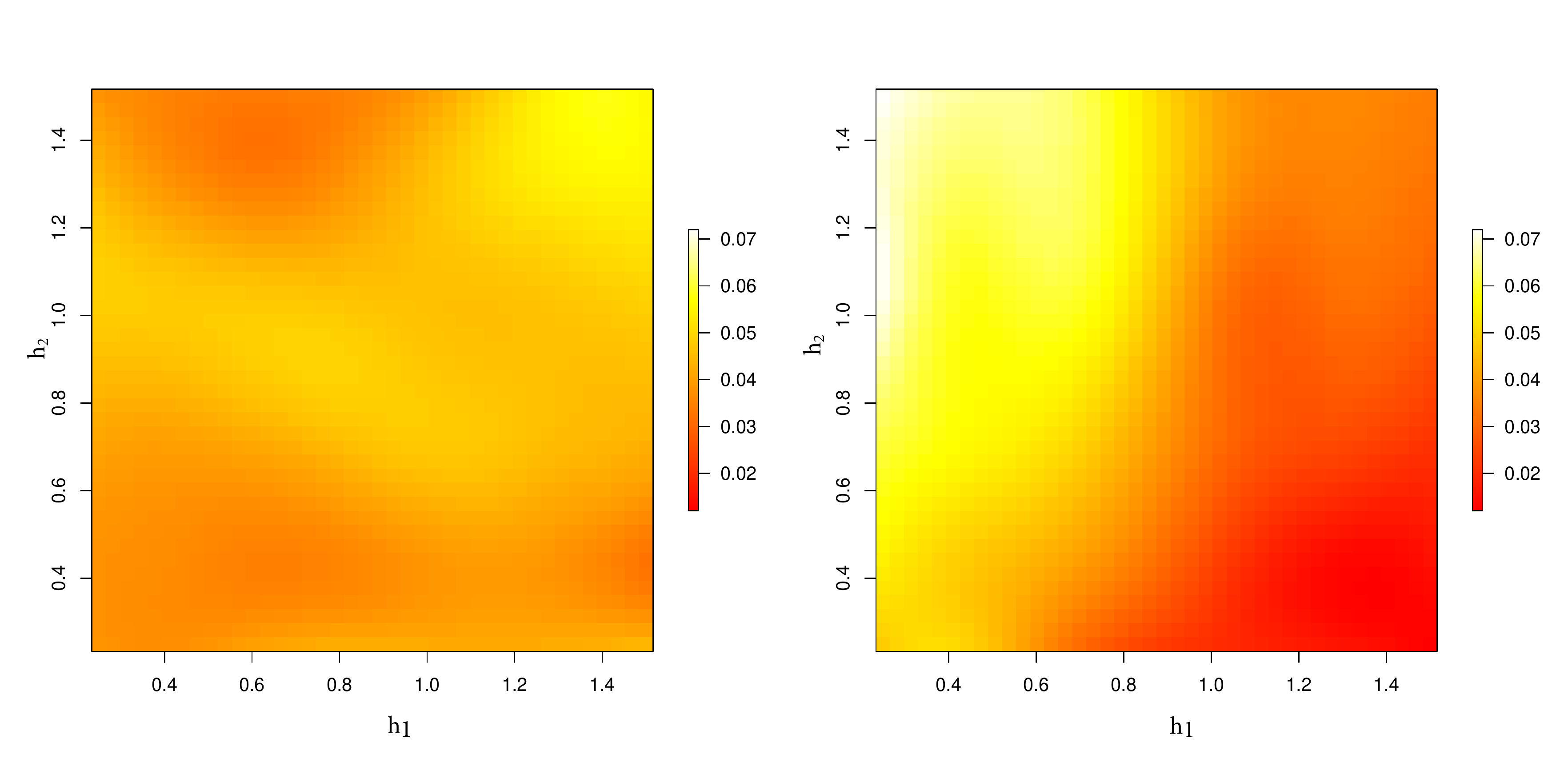}
	\caption{Proportions of rejections of the null hypothesis ($c=0$), for $\alpha=0.05$,  considering $c_0=0.04$, $\sigma^2=0.16$, $a_e=0.6$ and $n=400$, using PB (left) and CNPB (right), for several values of $h_1$ and $h_2$}
	\label{fig::simu_trace}
\end{figure*}

This section contains additional simulations, similar to those presented before, but  considering diagonal bandwidths with different elements, $\mathbf{H}=\text{diag}(h_1,h_2)$. Values of  $n=400$, $\sigma^2=0.16$, $c_0=0.04$  and $a_e=0.6$ were fixed.  The proportions of rejections (under the  null hypothesis, $c=0$) for different combinations of $h_1$ and $h_2$, and $\alpha=0.05$, are plotted in Fig.~\ref{fig::simu_trace}. Left panel of Fig.~\ref{fig::simu_trace} shows the results for PB and right panel for CNPB. Proportions of rejections for NPB are omitted due to its deficient calibration.  For this scenario, it can be observed that for PB  there are not relevant differences in terms of rejection proportions if $\mathbf{H}=\text{diag}(h,h)$ or $\mathbf{H}=\text{diag}(h_1,h_2)$ (with $h_1\neq h_2$) are considered.  Regarding CNPB, the use of a more general  bandwidth matrix does not provide better results with respect to using scalar bandwidth matrices. Although it is omitted here, similar conclusions can be obtained for alternative hypotheses ($c\neq 0$).

\section{Discussion}
\label{sec:discussion}
A goodness-of-fit test to assess a parametric trend surface for a geostatistical process is studied in this work. An exhaustive analysis of the behavior of the test considering different trends and dependence configurations is provided. The proposed test statistic measures the difference between a parametric and a nonparametric fits using an $L_2$-distance. An iterative least squares procedure has been used as a parametric trend estimator, whereas the multivariate Nadaraya--Watson estimator was employed as the nonparametric fit. The asymptotic distribution of the test statistic, under the null and under local alternatives, was derived considering the assumption of increasing-domain spatial asymptotics. 

For practical implementation, due to the slow convergence to the limit distribution, resampling methods were used to calibrate the test. Specifically, three
bootstrap procedures were designed and applied in practice:  PB, NPB and CNPB.   The CNPB resampling method avoids model selection and, therefore,  prevents against misspecification problems in the estimation
of the trend and dependence structure, unlike the PB approach. The CNPB also corrects the bias induced
by the use of the  residuals in the approximation of the dependence structure, using an iterative method, providing good results of the test under the null and alternative hypotheses. 
As it was pointed out in \cite{fernandez2014nonparametric}, a similar tool for bias adjustment could be included in the parametric
semivariogram estimation in the PB approach \citep[see][]{davison1997bootstrap}. However, this
way of proceeding would not avoid the misspecification problem in the parametric estimation of both the semivariogram and the  trend function.     

As usual in this type of problems, the performance of the goodness-of-fit test has been explored in a grid of different bandwidths to check how it is affected by the bandwidth choice. In the vast majority of scenarios considered in the simulation study, results obtained by the CNPB improve those achieved
by PB and NPB.  The use of non-scalar bandwidths has not provided better results for CNPB. The PB proposal works properly for calibration, but it shows a limited capacity to detect alternatives. On the other hand, although similar resampling methods to NPB have given good results when used in goodness-of-fit tests considering regression models with independent and univariate data, this is not the case in the geostatistical context. In this case,   the proportions of rejections under the null hypothesis are very large compared with the significance level considered, due to the underestimation of the variability of the process.

The three resampling approaches compared in this paper are based on computing the residuals from a pilot fit, estimating the corresponding covariance matrix of the errors and, finally, using a Cholesky decomposition to approximate a vector of \emph{independent} errors to generate bootstrap resamples. Other resampling procedures, such as the block bootstrap \citep[see][]{lahiri2013resampling}, could be used to calibrate the test. This method requires an appropriate partition of the observation region, unlike parametric and nonparametric bootstrap based methods. In addition, block bootstrap based approaches present difficulties  when the interest is focused on estimating the second-order
structure (dependence) of the process, which is often necessary to estimate properly the large-scale variability. These procedures fail to reproduce the variability of the process, thus leading to an underestimation of the semivariogram, possibly caused by the selection of the blocks. In \cite{castillo2019nonparametric},  parametric, corrected nonparametric and block bootstrap mechanisms were compared by checking  their performance in the approximation
of the bias and the variance of two variogram estimators. For inference on geostatistical processes and, particularly, on dependence structure estimation, the authors recommend the use of corrected nonparametric bootstrap methods. For these reasons, block bootstrap approaches were not employed in the present research.

The procedures used in the simulation study  were implemented in
the statistical environment \texttt{R} \citep{Rsoft}, using functions included in the   \texttt{npsp} and \texttt{geoR} packages  \citep{npsp, geor} to estimate the variogram and the spatial regression functions. In particular, the bias correction in CNPB  bootstrap algorithm is implemented in the  function \texttt{np.svariso.corr}  of the  \texttt{npsp} \texttt{R} package.



\section*{Acknowledgements}\label{acknowledgements}
The authors acknowledge the support from the Xunta de Galicia grant ED481A-2017/361 and the European Union (European Social Fund - ESF). This research has been partially supported by MINECO grants  MTM2016-76969-P and MTM2017-82724-R, and by the Xunta de Galicia (Grupo de Referencia Competitiva  ED431C-2017-38, and Centro de Investigación del SUG ED431G 2019/01), all of them through the ERDF.

\bibliography{bibib3}


\section*{Appendix. Proof of the main theorem}
In this appendix,  Theorem \ref{theorem}, under assumptions (A1)--(A8), is proved. The asymptotic distribution of the test statistic given in (\ref{eq:statistic}), comparing the nonparametric  and the smooth  parametric estimators, given in (\ref{eq:NW}) and (\ref{eq:NWpar}), respectively, using  an $L_2$-distance, is derived. 
\begin{proof}
	The test statistic (\ref{eq:statistic}) can be decomposed as
	\begin{eqnarray*}
		T_n&=&n\abs{\mathbf{H}}^{1/2}\int_{}^{}(\hat{m}^{NW}_{\mathbf{H}}(\mathbf s)-\hat{m}^{NW}_{\mathbf{H},\hat{\mathbf{\beta}}}(\mathbf s))^2w(\mathbf s)d\mathbf s\\&=&n\abs{\mathbf{H}}^{1/2}\int_{}^{}\left(\dfrac{\sum_{i=1}^n K_{\mathbf{H}}({\mathbf s_i-\mathbf s})Z_i}{\sum_{i=1}^n K_{\mathbf{H}}({\mathbf s_i-\mathbf s})}-\dfrac{\sum_{i=1}^n K_{\mathbf{H}}({\mathbf s_i-\mathbf s})m_{\hat{{\mathbf{\beta}}}}(\mathbf s_i)}{\sum_{i=1}^n K_{\mathbf{H}}({\mathbf s_i-\mathbf s})}\right)^2w(\mathbf s)d\mathbf s\\&=&n\abs{\mathbf{H}}^{1/2}\int_{}^{}\dfrac{\left[\sum_{i=1}^n K_{\mathbf{H}}\left({\mathbf s_i-\mathbf s}\right)\left(m(\mathbf s_i)+\varepsilon_i-m_{\hat{{\mathbf{\beta}}}}(\mathbf s_i)\right)\right]^2}{\left(\sum_{i=1}^n K_{\mathbf{H}}({\mathbf s_i-\mathbf s})\right)^2}w(\mathbf s)d\mathbf s.
	\end{eqnarray*}
	
	Now, taking into account that the trends considered are of the form $m=m_{{\mathbf{\beta}}_0}+ n^{-1/2}\abs{\mathbf{H}}^{-1/4}g$, one gets
	\begin{eqnarray*}\label{statistic}
		T_n&=&n\abs{\mathbf{H}}^{1/2}\int_{}^{}\dfrac{\left[\sum_{i=1}^n K_{\mathbf{H}}\left({\mathbf s_i-\mathbf s}\right)\left(m_{{\mathbf{\beta}}_0}(\mathbf s_i)+n^{-1/2}\abs{\mathbf{H}}^{-1/4}g(\mathbf s_i)+\varepsilon_i-m_{\hat{{\mathbf{\beta}}}}(\mathbf s_i)\right)\right]^2}{\left(\sum_{i=1}^n K_{\mathbf{H}}({\mathbf s_i-\mathbf s})\right)^2}w(\mathbf s)d\mathbf s
		\\&=&n\abs{\mathbf{H}}^{1/2}\int_{}^{}\left(I_1(\mathbf s)+I_2(\mathbf s)+I_3(\mathbf s)\right)^2w(\mathbf s)d\mathbf s,
	\end{eqnarray*}
	where
	\begin{eqnarray*}
		I_1(\mathbf s)&=&\dfrac{\sum_{i=1}^n K_{\mathbf{H}}\left({\mathbf s_i-\mathbf s}\right)\left(m_{{\mathbf{\beta}}_0}(\mathbf s_i)-m_{\hat{{\mathbf{\beta}}}}(\mathbf s_i)\right)}{\sum_{i=1}^n K_{\mathbf{H}}({\mathbf s_i-\mathbf s})},\\
		I_2(\mathbf s)&=&\dfrac{\sum_{i=1}^n K_{\mathbf{H}}\left({\mathbf s_i-\mathbf s}\right)n^{-1/2}\abs{\mathbf{H}}^{-1/4}g(\mathbf s_i)}{\sum_{i=1}^n K_{\mathbf{H}}({\mathbf s_i-\mathbf s})},\\
		I_3(\mathbf s)&=&\dfrac{\sum_{i=1}^n K_{\mathbf{H}}\left({\mathbf s_i-\mathbf s}\right)\varepsilon_i}{\sum_{i=1}^n K_{\mathbf{H}}({\mathbf s_i-\mathbf s})}.
	\end{eqnarray*}
	Under assumptions (A1)--(A2) and (A6), and given that the difference  $m_{\hat{{\mathbf{\beta}}}}(\mathbf s)-m_{{{\mathbf{\beta}}}_0}(\mathbf s)=O_p(n^{-1/2})$, it is obtained that
	\begin{eqnarray*}
		n\abs{\mathbf{H}}^{1/2}\int_{}^{}I_1^2(\mathbf s)w(\mathbf s)d\mathbf s&=&n\abs{\mathbf{H}}^{1/2}\int_{}^{}\left[\dfrac{\sum_{i=1}^n K_{\mathbf{H}}\left({\mathbf s_i-\mathbf s}\right)\left(m_{{\mathbf{\beta}}_0}(\mathbf s_i)-m_{\hat{{\mathbf{\beta}}}}(\mathbf s_i)\right)}{\sum_{i=1}^n K_{\mathbf{H}}({\mathbf s_i-\mathbf s})}\right]^2w(\mathbf s)d\mathbf s\nonumber\\&=&O_p(\abs{\mathbf{H}}^{1/2}).
	\end{eqnarray*}	
	
	For the  term $I_2(\mathbf s)$, using the assumption (A2), it follows that
	\begin{eqnarray}\label{b1H}
	n\abs{\mathbf{H}}^{1/2}\int_{}^{}I_2^2(\mathbf s)w(\mathbf s)d\mathbf s&=&n\abs{\mathbf{H}}^{1/2}\int_{}^{}\left(\dfrac{\sum_{i=1}^n K_{\mathbf{H}}\left({\mathbf s_i-\mathbf s}\right)n^{-1/2}\abs{\mathbf{H}}^{-1/4}g(\mathbf s_i)}{\sum_{i=1}^n K_{\mathbf{H}}({\mathbf s_i-\mathbf s})}\right)^2w(\mathbf s)d\mathbf s\nonumber\\&=& n\abs{\mathbf{H}}^{1/2}n^{-1}\abs{\mathbf{H}}^{-1/2}\int_{}^{} \left({K}_{\mathbf{H}}\ast g(\mathbf s)\right)^2w(\mathbf s)d\mathbf s\nonumber\\&=& \int_{}^{} \left({K}_{\mathbf{H}}\ast g(\mathbf s)\right)^2w(\mathbf s)d\mathbf s,
	\end{eqnarray}
	which corresponds to $b_{1\mathbf{H}}$ in Theorem 1.  Finally, $I_3(\mathbf s)$ (associated with the error component) can be decomposed as
	\begin{eqnarray*}\label{decom}
		n\abs{\mathbf{H}}^{1/2}\int_{}^{}I_3^2(\mathbf s)w(\mathbf s)d\mathbf s&=&n\abs{\mathbf{H}}^{1/2}\int_{}^{}\left(\dfrac{\sum_{i=1}^n K_{\mathbf{H}}\left({\mathbf s_i-\mathbf s}\right)\varepsilon_i}{\sum_{i=1}^n K_{\mathbf{H}}({\mathbf s_i-\mathbf s})}\right)^2w(\mathbf s)d\mathbf s\nonumber\\&=& n\abs{\mathbf{H}}^{1/2}\int_{}^{}\dfrac{\sum_{i=1}^n K^2_{\mathbf{H}}\left({\mathbf s_i-\mathbf s}\right)\varepsilon^2_i}{\left(\sum_{i=1}^n K_{\mathbf{H}}({\mathbf s_i-\mathbf s})\right)^2}w(\mathbf s)d\mathbf s\nonumber\\&+& n\abs{\mathbf{H}}^{1/2}\int_{}^{}\dfrac{\sum_{i\neq j} K_{\mathbf{H}}\left({\mathbf s_i-\mathbf s}\right)K_{\mathbf{H}}\left({\mathbf s_j-\mathbf s}\right)\varepsilon_i\varepsilon_j}{\left(\sum_{i=1}^n K_{\mathbf{H}}({\mathbf s_i-\mathbf s})\right)^2}w(\mathbf s)d\mathbf s\nonumber\\&=&I_{31}+I_{32}.
	\end{eqnarray*}

	Close expressions of $I_{31}$ and $I_{32}$ can be obtained computing the expectation and the variance of these terms. Under assumption (A6), it can be proved that
	\begin{eqnarray}\label{expI31}
	\mathbb{E}(\abs{\mathbf{H}}^{1/2}I_{31})\nonumber&=&\mathbb{E}\left[n\abs{\mathbf{H}}\int_{}^{}\dfrac{\sum_{i=1}^n K^2_{\mathbf{H}}\left({\mathbf s_i-\mathbf s}\right)\varepsilon^2_i}{\left(\sum_{i=1}^n K_{\mathbf{H}}({\mathbf s_i-\mathbf s})\right)^2}w(\mathbf s)d\mathbf s\right]\nonumber\\&=& {}\sigma^2K^{(2)}(0)\int_{}^{}w(\mathbf s)d\mathbf s\cdot\left(1+o(1)\right).
	\end{eqnarray} 	
	Similarly,  using  assumptions (A3), (A6) and (A7),  it can be obtained that 	
	\begin{eqnarray*}
		\mbox{Var}(\abs{\mathbf{H}}^{1/2}I_{31})&=&{\mbox{Var}}\left[n\abs{\mathbf{H}}\int_{}^{}\dfrac{\sum_{i=1}^n K^2_\mathbf{H}\left({\mathbf s_i-\mathbf s}\right)\varepsilon^2_i}{\left(\sum_{i=1}^n K_{\mathbf{H}}({\mathbf s_i-\mathbf s})\right)^2}w(\mathbf s)d\mathbf s\right]\\&=&2n^2\abs{\mathbf{H}}^2\sum_{i=1}^n\sum_{j=1}^n\int_{}^{}\int_{}^{}\dfrac{K^2_\mathbf{H}\left({\mathbf s_i-\mathbf s}\right)K^2_\mathbf{H}\left({\mathbf s_j-\mathbf{t}}\right)({\mbox{Cov}}(\varepsilon_i,\varepsilon_j))^2}{\left(\sum_{i=1}^n K_{\mathbf{H}}({\mathbf s_i-\mathbf s})\right)^2\left(\sum_{j=1}^n K_{\mathbf{H}}({\mathbf s_j-\mathbf{t}})\right)^2}w(\mathbf s)w(\mathbf{t})d\mathbf sd\mathbf{t}\\&=&2\sigma^4\abs{\mathbf{H}}\int_{}^{}\int_{}^{}\int_{}^{}\int_{}^{}K^2(\mathbf{v})K^2(\mathbf{z})w^2(\mathbf s)\rho_n^2(\mathbf{H}(\mathbf{v}-\mathbf{z}+\mathbf{u}))d\mathbf{v}d\mathbf{z}d\mathbf sd\mathbf{u}\cdot\left(1+o(1)\right).\\
	\end{eqnarray*}
	
	Let \begin{eqnarray*}
		j_n(\mathbf{v},\mathbf{u})&=&n\abs{\mathbf{H}}\int K^2(\mathbf{v})\rho^2_n(\mathbf{H}(\mathbf{v}-\mathbf{z}+\mathbf{u}))d\mathbf{z}.
	\end{eqnarray*}
	
	Notice that, using assumption (A3),
	\begin{eqnarray*}\abs{j_n(\mathbf{v},\mathbf{u})}&\le& K^2_M\left(n\abs{\mathbf{H}}\int \abs{\rho^2_n(\mathbf{H}(\mathbf{v}-\mathbf{z}+\mathbf{u}))}d\mathbf{z}\right)\\&\le&K^2_M\left(n\int \abs{\rho_n(\mathbf{t})}d\mathbf{t}\right)\\&\le&K_M^2\rho_{M},\end{eqnarray*}
	where $K_M\equiv\displaystyle \max_{\mathbf s}(K(\mathbf s))$ and $\rho_M\equiv\displaystyle \max_{\mathbf s}(\rho_n(\mathbf s))$, and using assumptions (A2), (A3), (A6) and (A8), one gets that
	\begin{eqnarray}\label{varI31}
	\mbox{Var}(I_{31})&=&o(1).
	\end{eqnarray}
	
	From (\ref{expI31}) and (\ref{varI31}) it follows that	
	\begin{equation}\label{b0H}
	I_{31}=\sigma^2 \abs{\mathbf{H}}^{-1/2}{}K^{(2)}(\mathbf{0})\int w(\mathbf s)d\mathbf s\cdot\left(1+o_p(1)\right).
	\end{equation}

	Now, consider the term   
	\begin{eqnarray*}       
		I_{32}&=&n\abs{\mathbf{H}}^{1/2}\int_{}^{}\dfrac{\sum_{i\neq j} K_{\mathbf{H}}\left({\mathbf s_i-\mathbf s}\right)K_{\mathbf{H}}\left({\mathbf s_j-\mathbf s}\right)\varepsilon_i\varepsilon_j}{\left(\sum_{i=1}^n K_{\mathbf{H}}({\mathbf s_i-\mathbf s})\right)^2}w(\mathbf s)d\mathbf s.\\
	\end{eqnarray*}
	
	Let 
	\begin{eqnarray*}
		\kappa_{ij}&=&
		n\abs{\mathbf{H}}^{1/2}\displaystyle\int_{}^{}\dfrac{K_{\mathbf{H}}\left({\mathbf s_i-\mathbf s}\right)K_{\mathbf{H}}\left({\mathbf s_j-\mathbf s}\right)}{\left(\sum_{i=1}^n K_{\mathbf{H}}({\mathbf s_i-\mathbf s})\right)^2}w(\mathbf s)d\mathbf s{\varepsilon}(\mathbf s_i){\varepsilon}(\mathbf s_j).  
	\end{eqnarray*}
	
	Thus,
	$$I_{32}=\sum_{i\neq j}\kappa_{ij},$$
	and this can be seen as a $U$-statistic with degenerate kernel. To obtain the asymptotic normality of $I_{32}$	we will apply the central limit theorem for reduced $U$-statistics under dependence given by \cite{kim2013central}.

	For this  term $I_{32}$ we have
	\begin{eqnarray*}
		\mathbb{E}\left(\abs{\mathbf{H}}^{1/2}I_{32}\right)&=&\mathbb{E}\left[n\abs{\mathbf{H}}\int_{}^{}\dfrac{\sum_{i\neq j} K_{\mathbf{H}}\left({\mathbf s_i-\mathbf s}\right)K_{\mathbf{H}}\left({\mathbf s_j-\mathbf s}\right)\varepsilon_i\varepsilon_j}{\left(\sum_{i=1}^n K_{\mathbf{H}}({\mathbf s_i-\mathbf s})\right)^2}w(\mathbf s)d\mathbf s \right]\\&=&n^{-1}\abs{\mathbf{H}}\sum_{i\neq j}\mbox{Cov}({\varepsilon}(\mathbf s_i),{\varepsilon}(\mathbf s_j))\displaystyle\int_{}^{}\dfrac{K_{\mathbf{H}}\left({\mathbf s_i-\mathbf s}\right)K_{\mathbf{H}}\left({\mathbf s_j-\mathbf s}\right)}{\left(\sum_{i=1}^n K_{\mathbf{H}}({\mathbf s_i-\mathbf s})\right)^2}w(\mathbf s)d\mathbf s\cdot \left(1+o(1)\right) \\&=&(n-1)\abs{\mathbf{H}}\sigma^2\displaystyle\int_{}^{}\int_{}^{}\int_{}^{}{K\left(\mathbf{v}\right)K\left(\mathbf{z}\right)}\rho_n(\mathbf{H}(\mathbf{v}-\mathbf{z}))w(\mathbf s)d\mathbf{v}d\mathbf{z}d\mathbf s\cdot\left(1+o(1)\right).\end{eqnarray*}
	
	Under the assumptions (A4), (A5) and (A6)--(A8), as shown by \cite{liu2001kernel}, 
	$$\lim_{n\to\infty}n\abs{\mathbf{H}}\int K(\mathbf{v})K(\mathbf{z})\rho_n(\mathbf{H}(\mathbf{v}-\mathbf{z}))d\mathbf{v}d\mathbf{z}=K^{(2)}(0)\rho_c.$$
	
	It follows that
	\begin{equation}\label{biasT3}
	\mathbb{E}(\abs{\mathbf{H}}^{1/2}I_{32})=\sigma^2K^{(2)}(0)\rho_{c}\int_{}^{}w(\mathbf s)d\mathbf s\cdot\left(1+o(1)\right).
	\end{equation}

	Similarly, it can be obtained that the asymptotic variance of $I_{32}$ is 
	\begin{eqnarray}\label{vart3}V=\sigma^4 K^{(4)}(0)\int_{}^{} w^2(\mathbf s)d\mathbf s\left(1+\rho_{c}+2\rho^2_c\right).
	\end{eqnarray}
	
	The term $I_{32}$ converges in distribution to a normally distributed random variable with mean the second term of $b_{0\mathbf{H}}$ and variance $V$.
	
	In virtue of the Cauchy--Bunyakovsky--Schwarz inequality, the  cross terms in $T_n$ resulting from the products of $I_1$, $I_2$ and $I_3$ are all of small order. Therefore, combining the results given in the equations (\ref{b1H}) and (\ref{b0H}), and the asymptotic normality of $I_{32}$ (with its bias (\ref{biasT3}) and its variance (\ref{vart3})), it follows that  
	\begin{equation*}\label{T1}
	V^{-1/2}(T_n-b_{0\mathbf{H}}-b_{1\mathbf{H}})\to_{\mathcal{L}} N(0,1) \text{ as } n\to\infty.
	\end{equation*}
	where $Z$ is a standard normal variable and 
	\begin{eqnarray*}
		b_{0\mathbf{H}}&=& \abs{\mathbf{H}}^{-1/2}\sigma^2K^{(2)}(0)\int_{}^{}{w(\mathbf s)}d\mathbf s\left(1+\rho_{c}\right),\\
		b_{1\mathbf{H}}&=& \int_{}^{}\left({K}_{\mathbf{H}}\ast g(\mathbf s)\right)^2w(\mathbf s)d\mathbf s,\\
		V&=&\sigma^4 K^{(4)}(0)\int_{}^{} w^2(\mathbf s)d\mathbf s\left(1+\rho_{c}+2\rho^2_c\right). 	\end{eqnarray*}
\end{proof}
\end{document}